\documentclass[lettersize,journal]{IEEEtran}
\usepackage{amsmath,amsfonts}
\usepackage{algorithm}
\usepackage{array}
\usepackage[caption=false,font=normalsize,labelfont=sf,textfont=sf]{subfig}
\usepackage{textcomp}
\usepackage{stfloats}
\usepackage{url}
\usepackage{verbatim}
\usepackage{graphicx}
\usepackage{cite}
\usepackage{algpseudocode}
\usepackage{xcolor}
\usepackage{multirow}

\begin{document}

\title{Implicit Self-supervised Language Representation for Spoken Language Diarization}

\author{Jagabandhu Mishra, \IEEEmembership{Student Member, IEEE}, and S. R. Mahadeva Prasanna, \IEEEmembership{Senior Member, IEEE}
 \thanks{*Corresponding author(s). Jagabandhu~Mishra (jagabandhu.mishra.18@iitdh.ac.in) and S.R.M.~Prasanna (prasanna@iitdh.ac.in) are from the Department of Electrical Electronics and Communication Engineering, Indian Institute of Technology (IIT) Dharwad, India.}

}

\markboth{Journal of JSTSP Class Files,~Vol.~14, No.~8, August~2023}%
{Shell \MakeLowercase{\textit{et al.}}: A Sample Article Using IEEEtran.cls for IEEE Journals}


\maketitle

\begin{abstract}
In a code-switched (CS) scenario, the use of spoken language diarization (LD) as a pre-possessing system is essential. Further, the use of implicit frameworks is preferable over the explicit framework, as it can be easily adapted to deal with low/zero resource languages. Inspired by speaker diarization (SD) literature, three frameworks based on (1) fixed segmentation, (2) change point-based segmentation and (3) E2E are proposed to perform LD. The initial exploration with synthetic TTSF-LD dataset shows, using x-vector as implicit language representation with appropriate analysis window length ($N$) can able to achieve at per performance with explicit LD. The best implicit LD performance of $6.38$ in terms of Jaccard error rate (JER) is achieved by using the E2E framework. However, considering the E2E framework the performance of implicit LD degrades to $60.4$ while using with practical Microsoft CS (MSCS) dataset. The difference in performance is mostly due to the distributional difference between the monolingual segment duration of secondary language in the MSCS and TTSF-LD datasets. Moreover, to avoid segment smoothing, the smaller duration of the monolingual segment suggests the use of a small value of $N$. At the same time with small $N$, the x-vector representation is unable to capture the required language discrimination due to the acoustic similarity, as the same speaker is speaking both languages. Therefore, to resolve the issue a self-supervised implicit language representation is proposed in this study. In comparison with the x-vector representation, the proposed representation provides a relative improvement of $63.9\%$ and achieved a JER of $21.8$ using the E2E framework.   
\end{abstract}


\section{\label{sec:1} Introduction}

Spoken language diarization (LD) refers to the automatic extraction of monolingual segments from a given code-switched (CS) utterance. Till today, humans are the best language recognizer in the world~\cite{li2013spoken,muthusamy1994reviewing,nagarajan2004implicit}. In accordance with the language abstraction level, humans use pre-lexical information i.e. acoustic-phonetic, phonotactic, prosodic, and lexical information i.e.  words, and phrases to recognize the language~\cite{li2013spoken,ambikairajah2011language}. The majority of the available systems use acoustic-phonetic, phonotactic, and phoneme dynamics (combined to form syllable/sub-words) related information to recognize the language~\cite{li2013spoken,ambikairajah2011language}. The acoustic-phonetic information is extracted from the spectro-temporal representation and mostly captures the phoneme production mechanism~\cite{li2013spoken,ambikairajah2011language}. Similarly, the phonotactic information captures the language-specific phoneme distribution~\cite{li2013spoken}. Alternatively, with respect to language modeling the existing language recognition systems can be broadly categorized into (a) implicit and  (b) explicit systems. The implicit systems model the language information directly from the speech signal. On the other hand, the explicit systems model the language information through intermediate modeling of phonemes, Senones and tokens, etc. Both approaches have their own pros and cons. The intermediate modeling of the explicit approach requires transcribed speech data and also complicates the system design~\cite{nagarajan2004implicit}. In contrast, the use of an implicit approach poses a challenge for the modeling of language-specific long-term dynamics directly from speech signals~\cite{mishra2023challenges,mishra2021spoken,mishra2022issues}. However, the recently evolved deep learning frameworks like the recurrent neural network (RNN), time-delay neural network (TDNN), and transformer, etc. are able to show their success in the modeling of long-term dynamics~\cite{dawalatabad2020novel,sarma2018language,shah2020first,mishra2022importancelid}. Further, the perception study shown in~\cite{muthusamy1994reviewing}, shows humans can able to recognize the language, without knowing the grammatical details of the language.
Therefore motivates this work to explore implicit approaches for performing LD tasks.

  \begin{figure}
 \centering
\includegraphics[height= 130pt,width= 220pt]{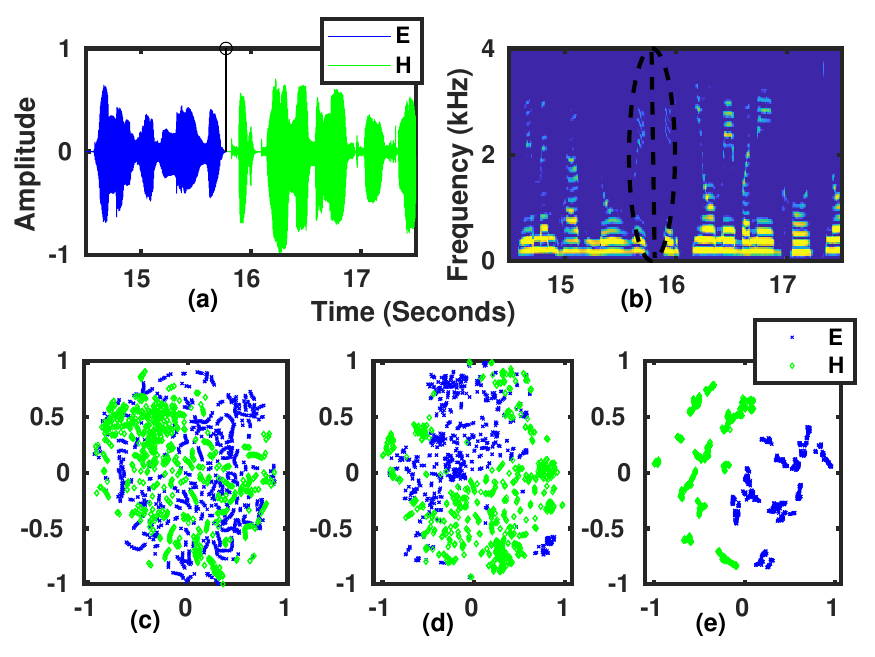}
 \caption{(a) Time domain representation of a Code-switched speech utterance, (b) spectrogram, (c) t-SNE distribution of the MFCC features, (d) W2V based ASR posterior and (e) x-vector representations, respectively.}
\vspace{-0.7 cm}
 \label{f1}
 \end{figure}

Specific to LD, mostly the CS utterances are spoken by a single speaker~\cite{mishra2023challenges,shah2020first}. In such a scenario, the phoneme production of secondary language may be biased toward the primary language and make language discrimination challenging at the acoustic-phonetic level. Fig.~\ref{f1}(a) and (b) shows the time domain and spectrogram representation of a CS utterance (Hindi-English). From both the time domain and spectrogram representation it is difficult to discriminate between the languages. Further, Fig.~\ref{f1}(c), (d), and (e) shows the  language-specific distribution of the two-dimensional t-SNE projection of $39$ dimensional voiced Mel frequency cepstral coefficients along with their velocity and acceleration (MFCC$+\Delta+\Delta\Delta$), posterior vectors extracted from wav2vec (W2V) finetuned Hindi and English model, and the TDNN based x-vector representations, respectively. The MFCC$+\Delta+\Delta\Delta$ features are extracted from speech signal by considering $20$ msec and $10$ msec as framesize and frameshift, respectively. The grapheme posterior vectors are extracted from the available trained automatic speech recognition (ASR) English ($32$ dimension) and Hindi ($67$ dimension) and models at~\cite{gupta2021clsril} and then concatenated to form $99$ dimension vectors with framesize and frameshift of $25$ and $20$ msec, respectively. The x-vectors are extracted from the implicitly trained x-vector framework by considering  framesize and  frameshift as $2000$ and $10$ msec, respectively. The figure shows that the overlap between the languages is more in the MFCC feature space. This is due to the similarity in the phoneme production of both primary and secondary language, as the secondary language phonemes are mostly produced by adapting the phoneme production system of the primary language. The overlap between the languages reduced significantly in the language-specific posterior and x-vector space. Here the language-specific posterior and x-vector represent the explicit and implicit system, respectively. Comparing Fig.~\ref{f1}(d) and (e), it can be observed that the language discrimination using the implicit approach is at par with the explicit approach. This observation justifies the feasibility of the development of the implicit LD system.

In literature there exist few attempts to perform LD and related tasks. The related tasks refer to CS point detection (CSD), CS utterance detection (CSUD), sub-utterance level language identification (SLID), etc.~\cite{shah2020first,mishra2022issues,liu2021end}. In some of the attempts, SLID task is termed LD as they predict language tags for each fixed duration segment within an utterance~\cite{liu2021end}. Mostly all the attempts try to capture either phonotactic information or the distribution they combine to form syllables and words using either an implicit or explicit approach. In ~\cite{lyu2013language} and ~\cite{barras2020vocapia}, the work attempts to perform SLID and CSUD tasks by considering both implicit and explicit approaches and observed that the explicit approach provides better performance than the implicit. The Gaussian posterior (GP) and i-vector approaches are the implicit approaches, whereas the phoneme posterior sequence (PS) extracted from the n-gram model is an explicit approach. In ~\cite{rallabandi2020detecting} and ~\cite{yilmaz2017language}, the work uses bottleneck features (BNF) extracted from the trained ASR as the language representation and latent features with variational Bayes encoder to perform CSD, CSUD, and LD tasks. In~\cite{shah2020first, rangan2020exploiting,krishna2020utterance} and ~\cite{liu2021end}, the works use deep learning architectures like the transformer, deepspeech2, and x-vector with deep clustering to implicitly model the language information for performing SLID and CSUD tasks. Though the work reported in ~\cite{lyu2013language} and ~\cite{barras2020vocapia} concludes the performance is better with explicit language modeling than the implicit, in ~\cite{krishna2020utterance} and ~\cite{liu2021end} for the CSUD task the performance of using implicit approach is at par with the performance achieved using the explicit approach. Furthermore, the advantage of using an implicit approach over an explicit approach is: (a) comparatively easier for implementation, (b) doesn't require intermediate phoneme/Senone modeling, (c) doesn't require transcribed speech data, and (d) easy to extend the technology for low resource and resource-scarce languages~\cite{nagarajan2004implicit}. Therefore, motivated to explore implicit approaches to performing the LD task.


Speaker diarization (SD) is a task, similar to LD and well explored in the literature~\cite{park2022review,tranter2006overview,moattar2012review}. Mostly the SD frameworks use the implicit approach to model the speaker information~\cite{park2022review,moattar2012review}.  Hence a close association study between SD and LD may help to develop the implicit LD systems. In the literature, the development of SD systems can be broadly categorized into three categories: (a) feature-based segmentation followed by clustering, (b) fixed segmentation and embedding-based (i/d/x-vector) clustering, and  (c) End-to-End based approach~\cite{park2022review}. 

The feature-based segmentation approach initially performs speaker segmentation, then the segments are used for the cluster initialization, followed by clustering to label the segments~\cite{moattar2012review}. In the fixed segmentation-based approach, the clusters are initialized by following a fixed duration segmentation approach and followed by clustering and label smoothing to obtain the segment-specific speaker labels~\cite{park2022review}. However, it is observed that, if the mono speaker segment duration has higher dynamic variation, the use of the smoothing window, even with optimal parameter setting affects the system performance~\cite{park2022review}. In contrast, the use of fixed-duration segmentation may also provide a performance trade-off in deciding upon the segment duration, due to the assumption that each segment should consist of speech samples from one speaker and should also have a duration sufficient enough to capture the speaker representation~\cite{park2022review}. Further, in~\cite{zhang2019fully}, proposed an end-to-end (E2E) framework to perform SD. The approach performs voice activity detection, speaker representation vector extraction, and clustering together using a single neural network by modifying the diarization task to a multi-label classification problem. In~\cite{fujita2019endpermuta}, the work eases out the training by proposing permutation-free training loss, and in~\cite{fujita2019end}, the BiLSTM and deep clustering are replaced with the self-attention-based framework. Though the E2E provides better performance than the traditional approach, the architecture poses a limit on the maximum number of speakers and also overfits with the training data distribution~\cite{park2022review}. This shows each framework has its own pros and cons. Therefore, inspired by the same the aim is to explore the LD using three frameworks: (a) fixed segmentation followed by clustering, (b) change point-based segmentation followed by clustering, and (c) E2E framework.

The study reported in ~\cite{mishra2021spoken} shows, to implicitly derive language representation requires a larger analysis window duration and a priori language information. However, the study reported in~\cite{shah2020first,mishra2022issues}, shows the duration of the secondary language segment is very small. In such a scenario, the hypothesis is the secondary language segments will be smoothed out and leads to a reduction in diarization performance. Keeping this in mind, initial frameworks are proposed and tuned up using a synthetically generated dataset. The dataset is synthetically generated such that the duration of primary and secondary language is approximately the same and the median is more than $4$ seconds. After that, the systems are evaluated in the practical dataset. The generative training of a restricted Boltzmann machine (RBM) based deep belief network (DBN) with unlabeled data, and fine-tuning of the same with a small amount of labeled data, were shown to be successes in classification tasks, with limited and imbalance training data~\cite{bengio2007greedy}. Motivating by the same, the hypothesis here is, the self-supervised training of the wav2vec (W2V) framework with unlabeled data, and fine-tuning of the framework with the limited imbalanced training data may provide better language representation with a small analysis window. Therefore the same is evaluated by extracting implicit language representations in all three frameworks. 


The rest of the work is organized as follows: Section~\ref{data} describes the details about the databases used in this study. 
 Section~\ref{LD_syn} describes the proposed implicit LD frameworks and also compares the performance with the explicit LD using the synthetically generated dataset. The proposed implicit frameworks are evaluated in the practical dataset and discussed in Section~\ref{LD_prac}. Section~\ref{SSL}, discusses the exploration of extracting self-supervised implicit representation and performing LD using them. In Section~\ref{dis} the initial hypothesis and obtained results are discussed. Finally, the conclusion and future direction are discussed in Section~\ref{con}.


\section{Database details}
\label{data}

This section provides a brief description of the databases used in this study. Initially, the systems are developed using a synthetic dataset that is generated from the Indian Institute of Technology Madras text-to-speech corpus (IITM-TTS)~\cite{baby2016resources}. After that, a practical CS dataset distributed by Microsoft is used to perform the LD experiments~\cite{shah2020first}.

The IITM-TTS dataset consists of recordings from each person in two languages (a) native and (b) English. Specific to a native language one male and one female speaker's speech utterances are available. The dataset consists of utterances from $13$ Indian languages and English. This study considers a native Hindi female speaker's utterances to generate data for the LD study. Similarly, the Hindi speaker's English utterances and an Assamese native female speaker's English utterances are used to generate data for the SD study. From both the selected partitions, the first $5$ hours of data per language/speaker is kept for training, and the rest is used for testing. The test partition is then used to synthetically generate the CS test utterances. The training partition is directly used to train the models like x-vector, whereas for W2V fine-tuning and E2E model training the synthetically CS utterances are generated from the training partition.  For both speaker and language, $10000$ and $4000$, training and testing CS utterances are synthetically generated. The mean of the monolingual segment duration of the Hindi and English languages are $6.5$ and $5.2$ seconds, respectively. Similarly, the mean of the mono-speaker segment duration of the generated two-speakers utterance is $5.19$ and $4.86$ seconds, respectively. The generated dataset for LD and SD study is called TTSF-LD and TTSF-SD, respectively.

The Microsoft code-switched (MSCS) corpus consists of CS utterances from three language pairs: (a) Gujarati-English (GUE), (b) Tamil-English (TAE), and (c) Telugu-English (TEE). Along with the utterances the language tags are also available for each segment of duration $200$ msec. The dataset consists of two partitions i.e. training and development. The training and development set consists of approximately $16$ and  $2$ hours of data from each language pair, respectively.  The mean segment duration of the primary and secondary languages is approximately $1.5$ and $0.5$ seconds, respectively. Further details of the dataset can be found at~\cite{shah2020first}.

\section{Spoken Language Diarization with TTSF-LD dataset}
\label{LD_syn}

In this section, LD is performed using the synthetically generated TTSF-LD dataset. Further, a comparison between implicit and explicit approaches is also performed. The details are discussed in the following subsections.

\subsection{Diarization with Implicit x-vector Representation }
\begin{figure}
 \centering
\includegraphics[height= 180pt,width= 240pt]{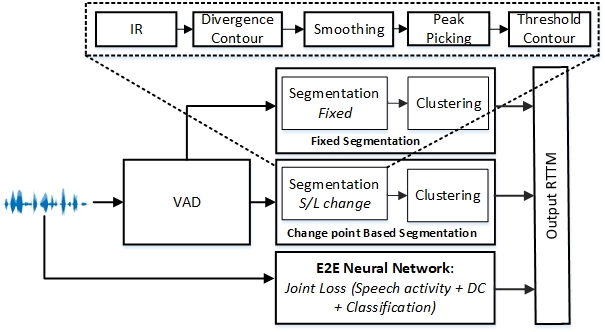}
 \caption{Block diagram of implicit diarization framework. VAD: voice activity detection, IR: implicit representation, and DC: deep clustering.}
 \label{f5}
 \end{figure}

Our earlier study reported in~\cite{mishra2021spoken} suggests the x-vector is a better implicit representation of language with an analysis window length of $N=200$. Hence in this study, the x-vector is used as a representation of language to perform LD. Further motivated by the SD literature three frameworks are proposed to perform LD, these are (1) diarization with fixed segmentation, (2) diarization with change point inspired segmentation, and (3) end-to-end diarization. The overview of each framework is depicted in Figure~\ref{f5}. Initially, to benchmark the framework, SD is performed using the x-vector as speaker representation with the synthetically generated multispeaker utterances (TTSF-SD). The details about the x-vector extraction and proposed frameworks are discussed in the following subsections.

\subsubsection{Training of x-vector architecture}
\begin{figure}
 \centering
\includegraphics[height= 90pt,width= 240pt]{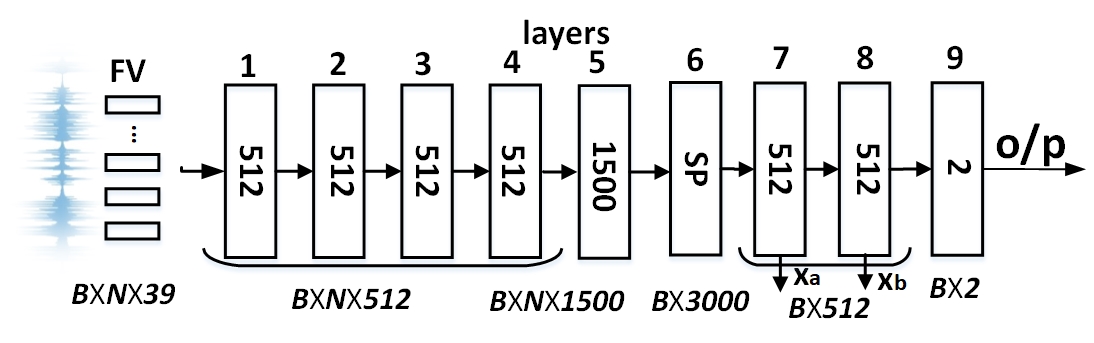}
 \caption{Block diagram of x-vector architecture. FV: MFCC feature vector, B: batch size, N: analysis window length, and $x_{a}$/$x_{b}$: x-vector.}
 \label{f6}
 \end{figure}

The language/speaker-specific training set data of TTSF-LD and TTSF-SD are used to train the x-vector architecture for language and speaker, respectively. The block diagram with the signal flow of the x-vector architecture is depicted in Figure~\ref{f6}. The architecture has $5$ hidden layers, $4$ with $512$ $1D$-CNN filters each, followed by a layer of $1500$ $1D$-CNN filters having kernel sizer of ($5$,$3$,$3$,$1$,$1$) and dilation of ($1$,$2$,$3$,$1$,$1$), respectively, working at the frame level. The output of the $5^{th}$ layer is statistically pooled to obtain the segment-level representation. Then  another $2$ hidden layers of $512$ linear neurons each and a classification layer of $2$ linear neurons work at the segment level. Each layer except the output is batch normalized and used leaky Relu as the activation function. Softmax is used as the activation function for the output layer. For input to the architecture, $39$ dimensional voiced MFCC features with their velocity and acceleration  coefficients are used. The MFCC features are extracted by considering $0.02$ and $0.01$ seconds as the framesize and frameshift, respectively. The voiced frames are decided by using an energy-based VAD.

The training of the speaker model doesn't require dropout and L2 norm, whereas for language both dropout and L2 norm are used for better convergence. In the language model, the dropout of $0.2$ is used in the $2$,$3$,$4$ and $7^{th}$ layer, respectively. For both speaker and language learning rate scheduler is used with the ADAM optimizer. To train the framework speech brain and PyTorch repositories are used. Inspired by our earlier study reported in~\cite{ravanelli2021speechbrain}, the speaker model is trained using $N=50$, and the language model with $N=50$ and $200$. The training is done for $100$ epochs for both the speaker and language model. By observing the validation loss and accuracy the model corresponding to epoch $11^{th}$, $15^{th}$ and $10^{th}$ are considered for the speaker with $N=50$, the language with $N=200$ and $50$, respectively for inference.

\subsubsection{x-vector representation}
The x-vectors are extracted from both $7^{th}$ and $8^{th}$ layer of the architecture and termed as $x_{a}$ and $x_{b}$, respectively. Before applying the representations to the diarization framework, the discrimination ability of the representations is evaluated.  The class-specific x-vector representations from the training set are used to obtain the LDA/WCCN projection matrix and GPLDA classifier. The test set of the TTSF-LD and TTSF-SD are used to evaluate the discrimination ability. From the synthetically generated utterances, the voiced MFCC features are extracted, and then using the ground truth labels the class-specific features are pooled together. After that with respect to $N$, x-vectors are extracted from the trained architecture. Using the pool of x-vectors, randomly $2000$ within language/speaker (WL/WS) pairs are generated and $2000$ between language/speaker (BL/BS) pairs are generated. The extracted x-vectors are multiplied with the projection matrix and then used to compute the GPLDA scores. The EER is used as an objective to calibrate the distributional difference between the WL/WS and BL/BS GPLDA scores. 

\begin{figure}
 \centering
\includegraphics[height= 113pt,width= 200pt]{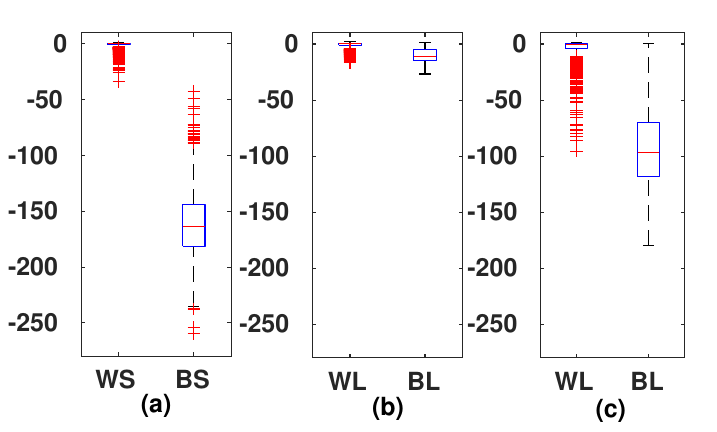}
 \caption{GPLDA score distribution of  x-vector representation between the trials of, (a) WS and BS with $N=50$ (EER$=0.001$), (b) WL and BL with $N=50$ (EER$=17$) and (c) with $N=200$ (EER$=3.6$), respectively.  }
 \label{f7}
 \end{figure}

The discrimination ability is verified for a given $N$ by considering $x_{a}$/$x_{b}$ as the x-vector with LDA and WCCN/ without LDA and WCCN. It is observed that for both tasks discrimination is better by considering $x_{a}$ as x-vector without LDA and WCCN. Hence all the tasks that are performed with TTSF-SD and TTSF-LD datasets are computed without LDA and WCCN and by considering $x_{a}$ as the x-vector representation. The obtained GPLDA score distributions between the WL/WS and BL/BS are depicted in Figure~\ref{f7}. 

It is observed that in the case of the speaker, with $N=50$ the WL and BL scores are well separated and provide EER of $0.001\%$. However, in the case of language with $N=50$, the WL and BL score distributions are more overlapped and provide EER of $17\%$. The overlap between the score distribution reduces by the increase in analysis window length and provides a best EER of $3.6\%$ at $N=200$. This shows for obtaining a better language representation implicitly requires a larger duration, as compared to the duration required for obtaining the speaker representation. Therefore for performing the diarization task, for the speaker, the x-vectors are extracted with $N=50$ and used as an implicit speaker representation. However, for comparing the effect of analysis window length in the LD task, the x-vectors are extracted by considering both $N=50$ and $N=200$.

\subsubsection{Diarization with fixed segmentation}
\begin{figure}
 \centering
\includegraphics[height= 100pt,width= 245pt]{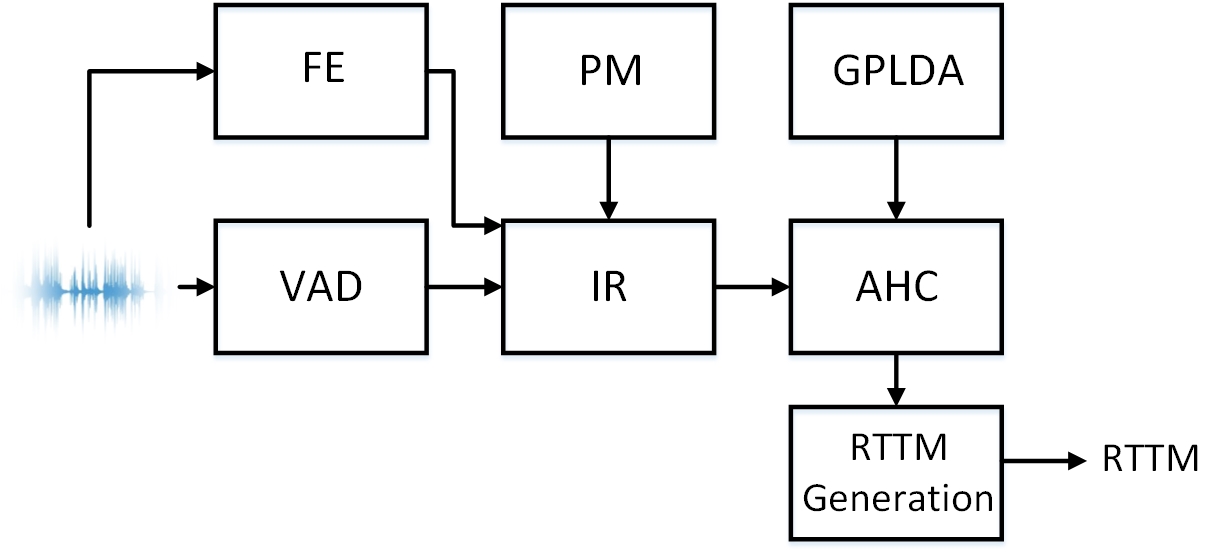}
 \caption{Diarization framework with fixed segmentation, VAD: voice activity detection, IR: implicit representation, PM: projection matrix, AHC: agglomerative hierarchical clustering.}
 \label{f8}
 \end{figure}

The fixed segmentation-based diarization framework is decided by motivating from the proposed SD frameworks in~\cite{park2022review,dawalatabad2020novel}. For performing the LD/SD, the given CS/multi-speaker test utterance is used to extract the $39$ dimensional MFCC features by considering $0.02$ and $0.01$ seconds as framesize and frameshift. At the same time, energy-based VAD is also performed for detecting the voiced frames. The start locations of the voiced frames are stored for further reference. The feature vectors belonging to the voiced frames are used to extract the implicit representation. In this case, x-vectors are extracted as an implicit representation using the trained x-vector extractor. The x-vectors are extracted by considering consecutive $N$ number of voiced feature vectors sequentially with a shift of one feature vector. The extracted x-vectors are then multiplied with the projection matrix to compensate for the intraclass variation and maximize the inter-class variation. In this case, as without the use of LDA/WCCN provides better language/speaker discrimination, an identity matrix of the dimension of $512 \times 512$ is used as a projection matrix (as per the dimension of x-vector i.e. $512$). The projected x-vectors are clustered using AHC by considering the maximum number of clusters as $2$ and the distance matrix as GPLDA with linkage as average distance. The cluster labels of each x-vector along with the stored voiced frame locations are used to generate the RTTM file of the given test utterance. The block diagram of the same is shown in Figure~\ref{f8}.

Mathematically, suppose the extracted feature vectors from the speech signal are represented as $\mathcal{F}=\{x_{1},x_{2},\ldots,x_{l}\}$, their corresponding frame energy as $\mathcal{E}=\{e_{1},e_{2},\ldots,e_{l}\}$, and their corresponding frames start location as  $\mathcal{P}=\{p_{1},p_{2},\ldots,p_{l}\}$. The voiced feature vectors can be computed as $\mathcal{F}_{v}=\{\mathcal{F}:x_{j}|e_{j} \geq 0.06 \times \Bar{e}\}$ and represented as $\{x^{v}_{1},x^{v}_{2},\ldots,x^{v}_{l'}\}$. Similarly, their locations as $\mathcal{P}_{v}=\{\mathcal{P}:p_{j}|e_{j} \geq 0.06 \times \Bar{e}\}$ and represented as $\{p^{v}_{1},p^{v}_{2},\ldots,p^{v}_{l'}\}$ . The representation vectors are extracted from the voiced frames as $r_{i}=\mathcal{X}(x^{v}_{i},x^{v}_{i+1},\ldots,x^{v}_{i+N-1})$ with a shift of one voiced frame and the set of representation vectors from an utterance can be represented as $\mathcal{R}=\{r_{1},r_{2},\ldots,r_{l''}\}$. $\mathcal{X}$ represents the trained x-vector architecture. Let's consider the obtained projection matrix (including both LDA and WCCN) from the training as $W$. The projected representation can be obtained as $\mathcal{R}_{p}=WR$ and represented as $\mathcal{R}_{p}=\{r^{p}_{1},r^{p}_{2},\ldots,r^{p}_{l''}\}$ . 

The representation vectors in $\mathcal{R}_{p}$ are then clustered through a bottom-up approach using the AHC algorithm. For clustering GPLDA is used as a distance matrix. The representation vectors are first mean-centered, whiten using the matrix obtained from training, and length normalized, after the vectors are represented as $\mathcal{R}_{\hat{p}}=\{r^{\hat{p}}_{1},r^{\hat{p}}_{2},\ldots,r^{\hat{p}}_{l''}\}$. The GPLDA distance between the vectors is computed using Eq.~\ref{hplda}, where $H_{1}$ and $H_{0}$ represent whether both the vectors are coming from the different classes and same class, respectively. By using Gaussian distribution, the equation can be further simplified to Eq.~\ref{gplda}, where $\Sigma$ represents the within-class variance, $S$ represents the between-class variance and the metrics are obtained during the training of GPLDA. 

\begin{equation}
    \label{hplda}
    D(r^{\hat{p}}_{i},r^{\hat{p}}_{j})=d_{ij}=\frac{p(r^{\hat{p}}_{i}|H_{1})p(r^{\hat{p}}_{j}|H_{1})}{p(r^{\hat{p}}_{i},r^{\hat{p}}_{j}|H_{0})}
\end{equation}

\begin{multline}
    \label{gplda}
d_{ij}=[r^{\hat{p}}_{i} ~ r^{\hat{p}}_{j}]^{T}   \begin{bmatrix}
\Sigma + SS^{T} & SS^{T} \\
SS^{T} & \Sigma + SS^{T}
\end{bmatrix}^{-1} [r^{\hat{p}}_{i} ~ r^{\hat{p}}_{j}]\\
-r^{\hat{p}^{T}}_{i}[\Sigma+ SS^{T}]^{-1}r^{\hat{p}}_{i}\\
-r^{\hat{p}^{T}}_{j}[\Sigma+ SS^{T}]^{-1}r^{\hat{p}}_{j}
\end{multline}

After the first merging, the distance of that cluster having more than one vector is obtained by averaging the individual vector distances and then used for further merging until reached two clusters. Suppose the obtained class levels are represented as $C=\{c_{1},c_{2},\ldots,c_{l''}\}$, and $c_{i} \in \{0,1,\ldots,C_{l}-1\}$, where $C_{l}$ is the number of classes. In this case, as we are dealing with two languages/speakers, hence $C_{l}=2$. The start locations of each representation vector i.e. $P_{rv}=\{p^{v}_{1},p^{v}_{2},\ldots,p^{v}_{l''}\}$ (discarding the $l''-l'$ frame locations from the end) are considered as the locations of each obtained class labels. For a given class level $c_{l}$, the obtained indices can be computed as $id=arg\{C:c_{i}|c_{i}==c_{l}\}$. The start locations of the given class label can be obtained as $s_{p}=\{P_{rv}:p_{j}^{v}|j \in [id_{1}, sid]\}$, where $sid$ is the discontinuities in the obtained indices and can be computed through first-order differentiation of the obtained indices i.e. $sid=\{id: id_{j}|j=arg\{(id_{i+1}-id_{i}) \neq 1\}+1,\forall i\}$. The end location of the segment can be identified as $e_{p}=\{P_{rv}:p_{j}^{v}|j \in [eid,id(end)]\}$, where  $eid=\{id: id_{j}|j=arg\{(id_{i+1}-id_{i}) \neq 1\},\forall i\}$. The duration of the segments present in the given class label can be computed as the difference between the start and end location i.e. $t=e_{p}-s_{p}$. After computing the start location and duration for each available class label, the sorting of start locations along with the duration and class label, will generate the predicted RTTM file for a given test utterance. 

\begin{figure}
 \centering
\includegraphics[height= 170pt,width= 245pt]{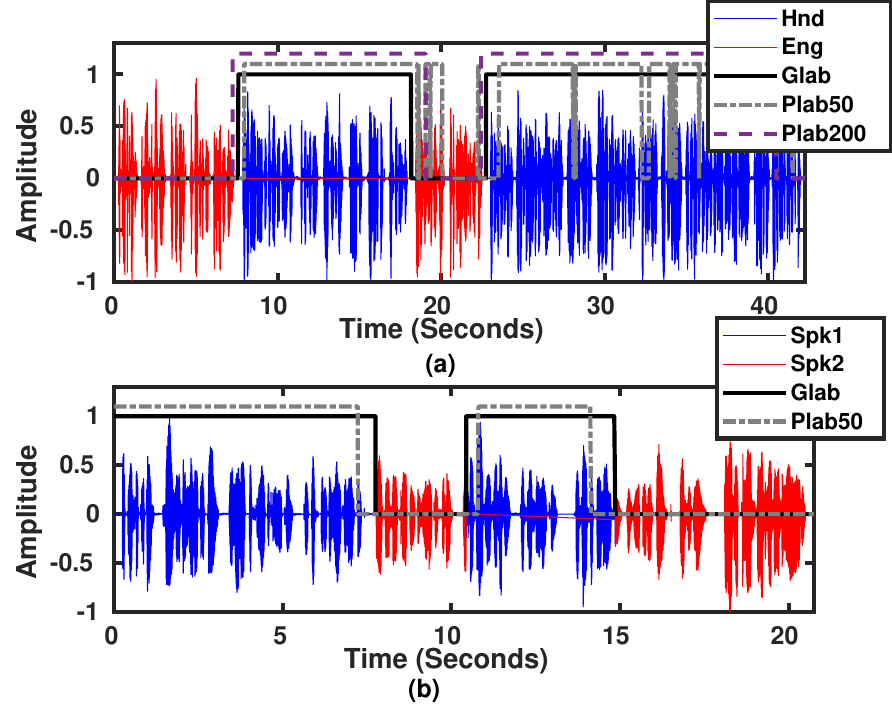}
 \caption{Predicted labeled segments, (a) LD,(b) SD, Glab: ground truth label, Plab50: predicted label with $N=50$ and Plab200: predicted label with $N=200$.}
 \label{f9}
 \end{figure}

Using the trained PLDA matrix and the extracted x-vectors from the test set of TTSF-LD with $N=50$, $N=200$, and TTSF-SD with $N=50$ the diarization is performed. The obtained results are tabulated in Table~\ref{FS-D}. The performance of the diarization is evaluated in terms of diarization error rate (DER) and Jaccard error rate (JER)~\cite{ryant2018first}. Throughout this work, the DER and JER are evaluated by considering the collar equal to zero. The tabulated DER and JER values are averaged across all the test utterances. The obtained DER and JER for the LD task with $N=50$ are $18.56$ and $30.19$ and with $N=200$ are $17.58$ and $29.39$, respectively. The difference in performance between $N=50$ and $200$ is because of the discrimination ability of the x-vector representation. The performance of the SD with $N=50$ is $10.03$ and $16.53$ in terms of DER and JER, respectively. This can be observed from Figure~\ref{f9}(a). In contrast to the label obtained with $N=200$, the label obtained with $N=50$ is fluctuating due to the inability of the x-vector representation.

In the case of SD, as observed from the discrimination ability the speaker representations are better than language (provides almost zero EER) and hence provide a better performance compare to language. Furthermore, even though the speaker representations provide almost zero EER, the DER of $10.03$ is due to the inability of the framework. Mostly, due to the confusion around the boundary as shown in Figure~\ref{f9}(b). Therefore required to further improve the framework. One way is to use the change point knowledge while clustering. Hence in the next section, a diarization framework is proposed to improve the diarization performance by including change point information while clustering.

\begin{table}
\centering
\caption{Performance of Language (TTSF-LD) and Speaker (TTSF-SD) diarization with fixed segmentation.}
\label{FS-D}
\begin{tabular}{|c|cc|c|}
\hline
    & \multicolumn{2}{c|}{TTSF-LD}       & TTSF-SD \\ \hline
N   & \multicolumn{1}{c|}{50}    & 200   & 50      \\ \hline
DER & \multicolumn{1}{c|}{18.56} & \textbf{17.58} & 10.03   \\ \hline
JER & \multicolumn{1}{c|}{30.19} & \textbf{29.39} & 16.53   \\ \hline
\end{tabular}
\end{table}

\subsubsection{Diarization with change point based segmentation}

\begin{figure}
 \centering
\includegraphics[height= 180pt,width= 245pt]{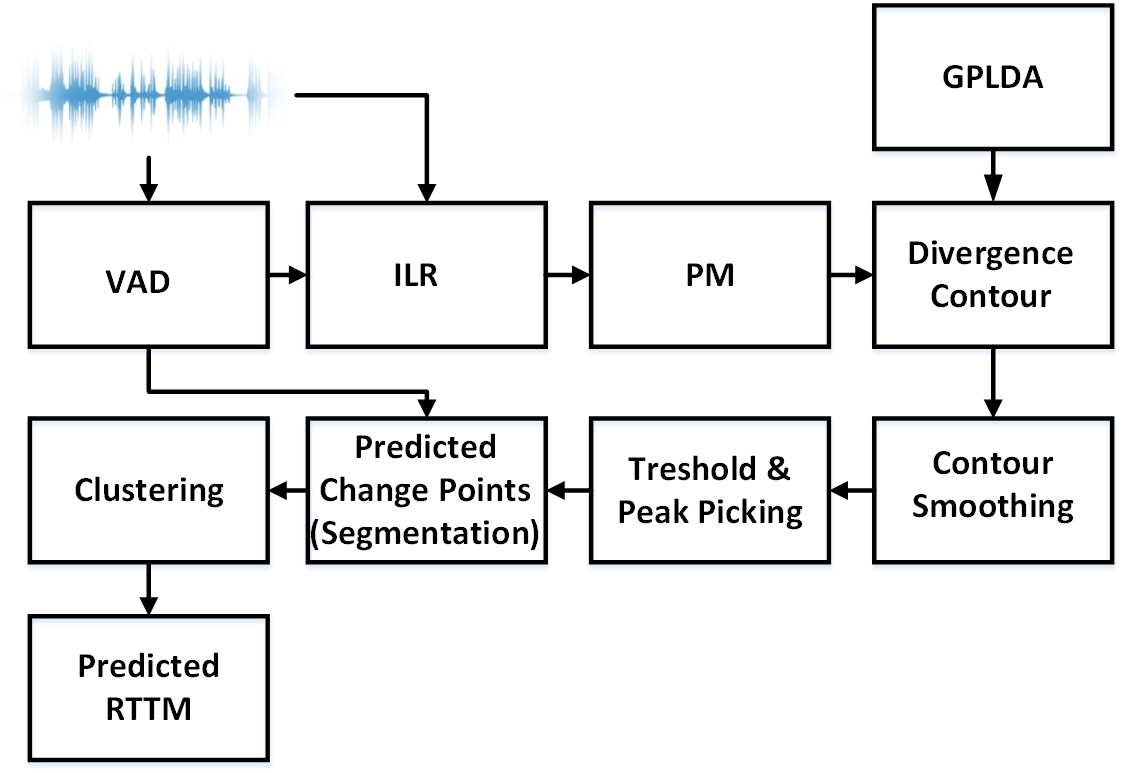}
 \caption{Implicit change point based LD framework, VAD: voice activity detection, ILR: implicit language representation.}
 \label{f3}
 \end{figure}

In the change point-based approach, change detection is initially performed, which confirms that each segment should consist of speech samples from one speaker/language depending on the task SD/LD. After that, from each segment, the implicit representation vector is extracted and then clustered using agglomerative hierarchical clustering (AHC). The clustering output is then further processed to obtain the RTTM files. The flow diagram of the same is depicted in Figure~\ref{f3}.

Suppose the voiced feature vectors and their corresponding locations of a given test utterance are represented as $\mathcal{F}_{v}=\{x_{1}^{v},x_{2}^{v},\ldots,x_{l'}^{v}\}$ and $\mathcal{P}_{v}=\{p_{1}^{v},p_{2}^{v},\ldots,p_{l'}^{v}\}$. Using the x-vector as the representation, the divergence contour is obtained using Eq.~\ref{dstf}, where $\mathcal{X}(.)$ represents the trained x-vector architecture, $W$ as the projection matrix and $\psi$ as the divergence computation function (in this case it is GPLDA). After obtaining the divergence contour, the contour is smoothed using a hamming window, i.e $D^{s}(i)=\sum_{k}D(k)h(i-k)$, where $h$ is the hamming window of length ($h_{l}$), and the $h_{l}$ is decided as $1/\delta$ times $N$. It is expected that whenever the change of class happens the smoothed contour archives its peak value. Hence a peak-picking algorithm is used to obtain the peak locations of smoothed divergence contour $D^{s}$. To avoid false alarms, the peak locations decided as the change locations are threshold by a minimum pick distance of $\gamma$ times $N$ and a threshold contour proposed in~\cite{lu2002speaker}. The threshold contour is obtained by using Eq.~\ref{kld_th}, where $N_{d}$ is the total length of $D^{s}$ and $\alpha$ is the threshold amplification factor. After getting the change locations of $D^{s}$ contour, the exact change locations can be obtained by mapping from $\mathcal{P}_{v}$ and can be represented as $\{c^{p}_{1},c^{p}_{2},\ldots, c^{p}_{M}\}$, where $M$ is the number of predicted change points. The midpoints of each predicted mono-language/speaker segment are computed as $c^{m}_{i}=\frac{c^{p}_{i+1}+c^{p}_{i}}{2}$. The midpoints are then used as a reference to obtain the x-vector representation from each predicted segment, i.e. $r_{i}=\mathcal{X}(x_{c^{m}_{i}-\frac{N}{2}+1},\ldots,x_{c^{m}_{i}+\frac{N}{2}})$. The representation vectors for each segment can be represented as $\mathcal{R}=\{r_{1},r_{2},\ldots,r_{M-1}\}$. The vectors are then projected using the LDA and WCCN projection matrix $W$, i.e. $\mathcal{R}^{p}=W\mathcal{R}$. The projected representation vectors are then clustered using AHC using GPLDA as the distance function. The clustered labels along with the segment start and end locations will be used to generate the predicted RTTM file for the given test utterance.

\begin{equation}
\label{dstf}
    D(i)=\psi(W\mathcal{X}(x^{v}_{i-N-1},\ldots,x^{v}_{i-1}),W\mathcal{X}(x^{v}_{i},\ldots,x^{v}_{i+N-1}))
\end{equation}

\begin{equation}
\label{kld_th}
 Th(i) =\alpha.\frac{1}{N_{d}}\sum_{n=0}^{N_{d}}D^{s}(i-n-1,i-n)
\end{equation}

The TTSF-LD and TTSF-SD dataset is used to perform the LD and SD using change point-inspired segmentation. For TTSF-LD, the diarization is performed with both $N=50$ and $200$, and for TTSF-SD,  with $N=50$. The hyperparameters ($\alpha,\delta ,\gamma$) are ($3.2$, $1.3$, $0.9$) used for TTSF-LD with $N=200$, and ($2.6$, $1.3$, $0.9$) with $N=50$, respectively. For TTSF-SD with $N=50$ the hyperparameters used are ($2.6$, $1.3$, $0.9$). The used hyperparameters are the optimal parameters that are obtained from the greedy search on the first $100$ test utterances in each case.

\begin{table}[]
\centering
\caption{Performance of Language and
Speaker diarization with change point based segmentation.}
\label{cp-D}
\fontsize{7pt}{7pt}\selectfont
\begin{tabular}{|ll|l|l|l|l|l|l|l|}
\hline
\multicolumn{2}{|l|}{}                         & N   & IDR   & MR    & FAR   & Dm   & DER            & JER            \\ \hline
\multicolumn{2}{|l|}{\multirow{2}{*}{TTSF-LD}} & 50  & 67.12 & 18.36 & 14.52 & 0.21 & 15.82          & 26.36          \\ \cline{3-9} 
\multicolumn{2}{|l|}{}                         & 200 & 87.01 & 4.41  & 8.84  & 0.28 & \textbf{11.16} & \textbf{20.61} \\ \hline
\multicolumn{2}{|l|}{TTSF-SD}                  & 50  & 92.27 & 3.96  & 3.76  & 0.03 & 6.84           & 13.42          \\ \hline
\end{tabular}
\end{table}

The obtained diarization along with the change detection performance is tabulated in Table~\ref{cp-D}. The change detection performance is also evaluated by following the event detection performance measures reported in~\cite{murty2008epoch,mishra2021spoken}. The used performance measures are identification rate (IDR), miss rate (MR), false acceptance rate (FAR), and mean deviation ( $D_m$). The measures don't depend on the tolerance window duration, instead calculated by observing the event activity in each region of interest (ROI) segment. The ROI segments are the duration between the mid-location of the consecutive ground-truth change points. The IDR defines the percentage of the segments for which exactly one change point is detected. FAR defines the percentage of the segments for which more than one change point is detected. MDR is the percentage of the segments for which no change point is
detected. $Dm$ is the mean deviation of the timing difference between the location of the detected and the ground truth change point.

The change detection performance of TTSF-LD with $N=50$ in terms of IDR, MR, and FAR are $67.12\%$, $18.36\%$, and $14.52\%$, respectively. The performance with $N=200$ is raised to $87.01\%$, $4.41\%$, and $8.84\%$, respectively. This is mostly due to the continuity nature of speech spectrum, even with language change. With increase in number of voiced frame the x-vector with tdnn able to learn the temporal dynamics of the language (may the inter and intra word level relation), and hence provides an improvement. However due to the large analysis window the mean delay ($D_{m}$) increases from $0.28$ seconds with $N=200$ compare to $0.21$ seconds with $N=50$. The abrupt change in the speech spectrum, during speaker change attributes in the change detection performance and provides the best performance with $N=50$ of $92.27\%$, $3.96\%$, $3.76\%$ and $0.03$ seconds in terms of IDR, MR, FAR and $D_m$, respectively. The similar trend is also observed with the diarization performance. The performance of diarization on TTSF-LD  with $N=50$ in terms of DER and JER is $15.82$ and $26.36$, respectively. The performance improves to $11.16$ and $20.61$ with $N=200$. As like change detection, the performance of SD is also better than the performance of LD and the obtained performance is $6.84$ and $13.42$ in terms of DER and JER, respectively.

 \begin{figure}
 \centering
\includegraphics[height= 170pt,width= 245pt]{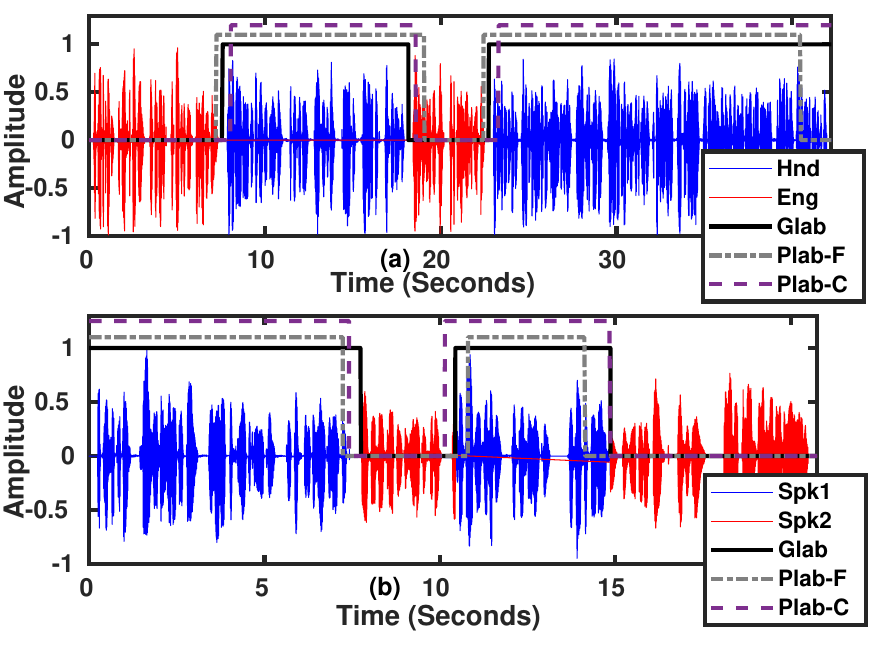}
 \caption{Predicted labeled segments, (a) LD, (b) SD, Glab: ground truth label, Plab-F: predicted label with fixed segmentation, and Plab-C: predicted label with change point inspired segmentation.}
 \label{f10}
 \end{figure}

In comparison to the fixed segmentation, it is observed that irrespective of the analysis window duration and speaker/language diarization, in all cases, the change point-inspired segmentation provides better performance. For LD the performance improved from $17.58$ and $29.39$ to $11.16$ and $20.61$, in terms of DER and JER, respectively. Similarly, the performance of SD also improved from $10.03$ and  $16.53$, to $6.84$ and $13.42$, respectively. This shows the significance of change point-inspired segmentation over the fixed. The same can be observed from Figure~\ref{f10}. From the figure, it can be observed that the problem with confusion around boundary region is reduced in both cases of language (Figure~\ref{f10}(a)) and speaker (Figure~\ref{f10}(a)) segment prediction. However, in the case of LD, there still exists a performance gap due to the discrimination ability of the x-vector representation. Motivating from the working principle of E2E diarization frameworks that the language/speaker representation and clustering can be optimized together, may further improve the performance by learning better discriminative representation. Therefore, the same is being explored for LD and discussed in the following subsection.

\subsubsection{End-to-end diarization}
\begin{figure}
 \centering
\includegraphics[height= 160pt,width= 180pt]{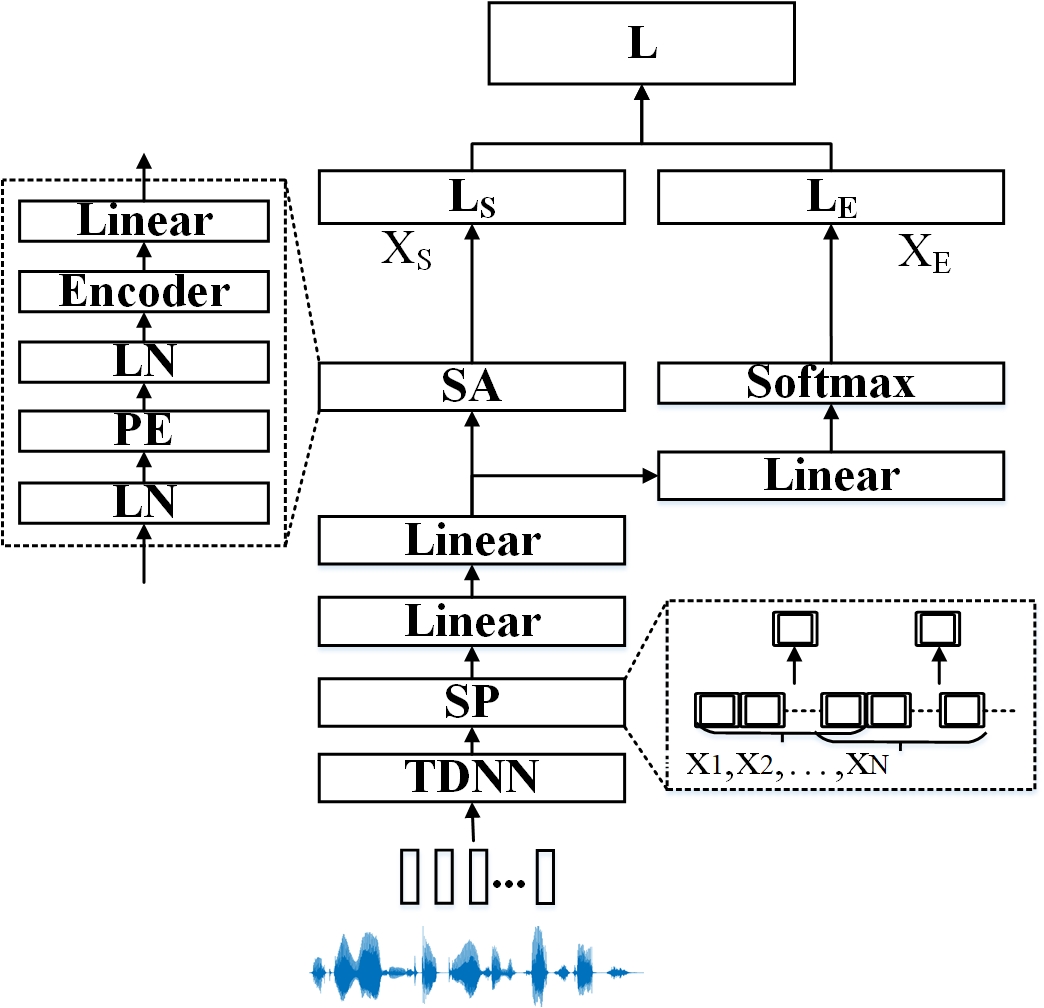}
 \caption{E2E diarization framework, SP: statistical pooling, SA: self-attention, LN: length normalization, and PE: positional encoding.}
 \label{f11}
 \end{figure}

This framework is inspired by the work reported in ~\cite{fujita2019end} and ~\cite{liu2021end}, for performing LD and SD. The beauty of the approach is, it performs VAD, clustering, and extraction of representation all together, through a joint loss. The block diagram of the framework is depicted in Figure~\ref{f11}. 

The TDNN block consists of four hidden layers, having $512$, $512$, $512$, and $1500$, numbers of $1D$-convolutional filters, respectively. The first two layers have a kernel size of $5$ and the next two have a kernel size of $1$. The dilation used for each hidden layer is $1$. The statistical pooling layer combines the $N$ frame-level representation with a shift of $N_{s}$ to a segment-level representation by concatenating the mean and diagonal covariance of the frame-level representations. The next two fully connected linear layers consist of $3000$, and $256$ neurons, respectively. For computing $X_E$, there have two fully connected layers having a number of neurons $256$ and $3$ (two classes and silence as another class), respectively. All the linear, (except the last)  and CNN layers use batch normalization and Relu as an activation function. The last layer having $3$ neurons use softmax as an activation function. Further, to compute $X_S$, the self-attention block consists of length normalization, followed by positional encoding, length normalization, four transformer encoders, and a linear fully connected layer. The transformer encoder uses $4$ head self-attention with $256$ dimensional key, $256$ dimensional value, and $2048$ fully connected neurons. The final linear layer consists of $3$ neurons having sigmoid as an activation function. The hierarchical structure of the architecture expects, initial TDNN layers to capture frame-level local information, the $X_E$ to capture the segment-level information  by tuning parameters through optimizing $L_E$ to obtain the better segment-level representation, and the $X_S$, global utterance level information through transformer layers by optimizing $L_S$ to predict best speaker/language sequence along with silence. The joint loss ($L$) is computed using equation~\ref{loss}, where $Y$ is the label sequence of a given utterance, CE(.) is the cross entropy loss and $w$ is the weighting parameter.

\begin{equation}
    \label{loss}
    L=wL_{S}+(1-w)L_{E}
\end{equation}
\begin{equation}
    \label{S_loss}
    L_{S}=CE(Y,X_{S})
\end{equation}
\begin{equation}
    \label{E_loss}
    L_{E}=CE(Y,X_{E})
\end{equation}

The framework is trained by using language/speaker along with silence sequence labels and extracted MFCC features from the training CS utterances. For a given utterance the label sequence is generated for each $200$ msec duration. The generated $10000$ training utterance from TTSF-LD and TTSF-SD dataset is used to train the E2E, LD, and SD architecture. The $N$ and $N_s$ considered for LD are $200$ and $20$ frames, respectively. For SD,  the $N$ and $N_s$ are considered $50$ and $20$ frames respectively. For each case, the E2E architecture is trained for $100$ epochs and the model parameters corresponding to the epoch that provides the best validation accuracy is chosen for testing.

During testing, the test set of TTSF-LD is used for LD, and TTSF-SD is used for SD, respectively. For a given test utterance the prediction levels are obtained from the trained E2E architecture at each $200$ msec. For obtaining the RTTM file, the language/speaker-specific segments are obtained by locating change boundaries in the predicted sequence. After obtaining the RTTM files for the test utterance the DER and JER  are computed by comparing them with the ground truth RTTM files. The obtained performance is tabulated in Table~\ref{e2e-d}. The performance is $5.81$ and $6.38$, respectively. It can be observed that, as expected the performance is better than the performance achieved using fixed and change point-inspired segmentation approaches. A similar trend is also being observed in SD. The performance of SD in terms of DER and JER is  $4.78$ and $4.99$, respectively.

\begin{table}[]
\centering
\caption{Performance of the E2E Language and
Speaker diarization.}
\label{e2e-d}
\begin{tabular}{|c|c|c|c|}
\hline
        & N   & DER           & JER           \\ \hline
TTSF-LD & 200 & \textbf{5.81} & \textbf{6.38} \\ \hline
TTSF-SD & 50  & 4.78          & 4.99          \\ \hline
\end{tabular}
\end{table}

\subsection{Explicit Spoken Language Diarization}
\label{ex_app}

This section aims to perform LD using the language representations obtained through the explicit approach. This study considers the grapheme posteriors of the fine-tuned wav2vec (W2V) models as the explicit language representation extractor.  As in this study, the explicit LD system is to be evaluated with Indian CS data i.e. TTSF-LD, it is assumed that the model pre-trained and fine-tuned with Indian data may provide better language representation. Hence, the W2V fine-tuned models available at~\cite{gupta2021clsril} are used here. The W2V framework is pre-trained with $10,000$ hours of speech data from $23$ Indian languages. The pre-trained model is fine-tuned using $4200$ and $700$ hours of transcribed speech data from the Hindi and Indian-English language separately. The English and Hindi model is trained considering $32$ and $67$ graphemes, respectively. The grapheme models of Hindi and English languages are represented as $G_{H}$ and $G_{E}$, respectively.  For a given utterance the posterior probabilities ($P(x|G_{H}^{i})$/$P(x|G_{E}^{i})$) are computed from both the English and Hindi models by considering $25$ msec and $20$ msec as frame length and  frameshift, respectively. The range of $i$ is $1 \leq i \leq P$, where $P$ is the number of grapheme posterior. After posterior computation, the Explicit W2V representation ($E^{W}$) is extracted by concatenating the obtained posteriors of English ($E^{W}_{E}$) and Hindi ($E^{W}_{H}$) fine-tuned model, i.e. $E^{W}=[ E^{W}_{E} ~ E^{W}_{H}]$, where $E^{W}_{E}$ and $E^{W}_{H}$ are $P(x|G_{E}^{i})$ and $P(x|G_{H}^{i})$, respectively. The extracted representations are then used with all three frameworks to perform the LD task. 


\subsubsection{Explicit Language Representation}

Initially, during training the monolingual speech utterances from TTSF-LD are used. The utterances are passed through a voice activity detector (VAD) and the locations of the energy frames are identified, by considering $20$ msec as framesize and frameshift. The explicit language representations ($E^{W}$) are extracted from the speech signal. The  $E^{W}$s' that belong to the voiced frames are only considered and the locations of the voiced frames are stored for future reference. The $E^{W}$ are used through a frame aggregation strategy with an analysis window length to obtain the aggregated vector.

A study is performed to obtain an optimal strategy to perform frame aggregation. The frame aggregation strategies compared here are, (1) i-vector with LDA and WCCN, (2) i-vector without LDA and WCCN, (3) mean pooling, (4) statistical (mean and standard deviation) pooling of $E^{W}$ representations with analysis window duration $N_{w}$.  As the explicit representations are extracted in each $20$ msec, the analysis window length ($N_{w}$) used to extract the explicit representation is half the analysis window length followed for the x-vector representations (i.e. $N_{w}=\frac{N}{2}$). Hence, for easy readability and comparison, even though the frame aggregation is performed with $N_{w}$, it will be represented throughout the work in terms of $N$ (i.e. $2 \times N_{w}$).


 \begin{figure}
 \centering
\includegraphics[height= 120pt,width= 240pt]{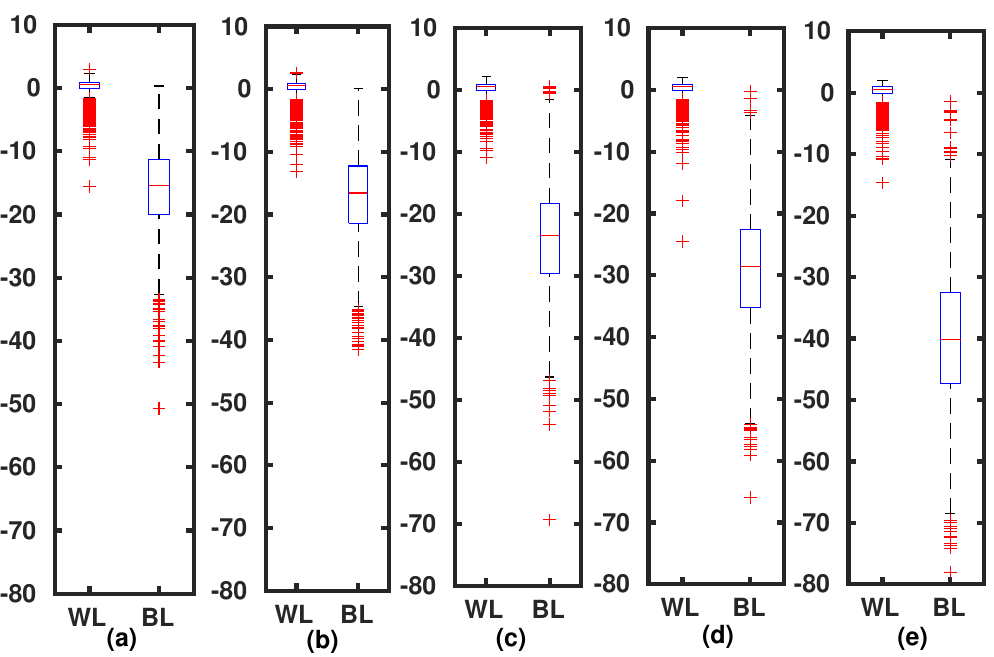}
 \caption{The GPLDA score distribution between WL and BL pairs, using frame aggregation as (a) i-vector with LDA, WCCN, (b) i-vector without LDA, WCCN, (c) mean pooling, (d) statistical pooling with $N=50$ and (e) statistical pooling with $N=100$. The obtained EERs are $2.05$, $1.7$, $0.8$, $0.25$ and $0.05$, respectively.}
 \label{f4}
 \end{figure}

Initially motivated by the i-vector framework, the voiced $E^{w}$ representations extracted from the training partition of TTSF-LD are used to train the universal background model (UBM). Using the UBM posteriors, sufficient statistics are obtained and the total variability (T) matrix is trained. After that from each utterance i-vectors are extracted from each analysis window by considering the length as $N$. The i-vectors are then used to, compute the LDA, and WCCN matrix and train the GPLDA classifier. 

The language discrimination ability of the frame aggregation strategies is tested by considering the test set of the TTSF-LD dataset. The obtained GPLDA score distributions are depicted in Figure~\ref{f4}. Figure~\ref{f4} (a) and (b) shows the WL and BL distributions of the i-vector strategy with LDA, WCCN, and without LDA, WCCN. It is observed that the overlapping between WL and BL reduces from Figure~\ref{f4}(a) to (b), with the EER of $2.05$ and $1.7$, respectively. Hence, for further strategies, LDA and WCCN are not performed. The overlap between the distributions is further reduced by using simple mean pooling and statistical pooling, which are shown in Figure~\ref{f4}(c) and (d), and the obtained EERs are $0.8$ and $0.25$, respectively. The $N$ value considered for Figure~\ref{f3}(a) to (d) is $50$ (approximately $500$ msec), and for i-vector computation the UBM is built with $32$ mixtures. Further, the $N$ value is increased from $50$ to $100$ to perform statistical polling, and the obtained WL and BL distributions are depicted in Figure~\ref{f4}(e). The obtained EER for the same is $0.05$. It is observed from the plots that the overlapping between the WL and BL GPLDA score distributions is least by using statistical pooling with $N$ equal to $100$. Though i-vectors have the ability to better aggregate the feature compared to the simple mean and statistical pooling, in this case, statistical pooling provides the best discrimination.  This may be due to the $E^{w}$ representations have already captured the temporal dynamics while pre-training and fine-tuning through the W2V framework, and may not require further complex projections to acquire temporal aggregation. Hence, for further processing, the statistical pooling is used as a frame aggregation strategy with $N$ equal to $100$.

\subsubsection{Performances of Explicit LD}

The explicit LD, system is developed using the TTSF-LD dataset. The training partition of the dataset is used to train the GPLDA. The test partition is used to evaluate the performance in all three LD frameworks. After extracting explicit representations, the same procedure as implicit is followed to obtain the LD performance. In the change point-based approach the tuned ($\alpha$, $\delta$, and $\gamma$) parameters are ($0.9$, $0.5$, and $0.5$), respectively. The fixed and change point segmentation-based frameworks are evaluated considering $N=100$ and a shift of a single frame (i.e. $20$ msec), whereas the E2E framework uses $N=100$ and a shift of $10$ (i.e. $200$ msec).  

The obtained performance is tabulated in Table~\ref{per_syn_exp}. The performance in terms of DER and JER for the fixed segmentation-based framework is $12.37$ and $20.74$. The performance is improved to $9.3$ and $17.23$ by using the change point-based segmentation framework. Furthermore, by using the E2E framework the performance further improved to $10.7$ in terms of JER, however, the performance of $11.37$ in terms of DER is at par with the performance change point based segmentation framework.




\begin{table}[]
\centering
\caption{Performance of explicit LD on the synthetically generated TTSF-LD dataset, FS: fixed segmentation}
\label{per_syn_exp}
\begin{tabular}{|c|ccc|}
\hline
\multicolumn{1}{|l|}{} & \multicolumn{3}{c|}{Explicit}                                            \\ \hline
N                      & \multicolumn{3}{c|}{100}                                                 \\ \hline
\multicolumn{1}{|l|}{} & \multicolumn{1}{c|}{FS}    & \multicolumn{1}{c|}{CPS}   & E2E            \\ \hline
DER                    & \multicolumn{1}{c|}{12.37} & \multicolumn{1}{c|}{9.3}   & \textbf{11.37} \\ \hline
JER                    & \multicolumn{1}{c|}{20.74} & \multicolumn{1}{c|}{17.23} & \textbf{10.7} \\ \hline
\end{tabular}
\end{table}



\section{Diarization with practical CS utterances}
\label{LD_prac}
The MSCS dataset consists of data from three language pairs GUE, TAE, and TEE used as a practical data set to perform LD study. The dataset consists of training and development partitions. In each set, the ratio of primary to secondary language data duration is $4:1$. Each utterances have a language tag available at every $200$ msec, using them ground truth RTTM files are generated for each utterance in the development set. All three diarization frameworks are trained using the training set and evaluated using the development set as test set data. The details of the training procedure and the obtained performances are discussed in the following subsections.

\subsubsection{Implicit LD with Fixed and change point inspired segmentation}

\begin{figure}
 \centering
\includegraphics[height= 120pt,width= 220pt]{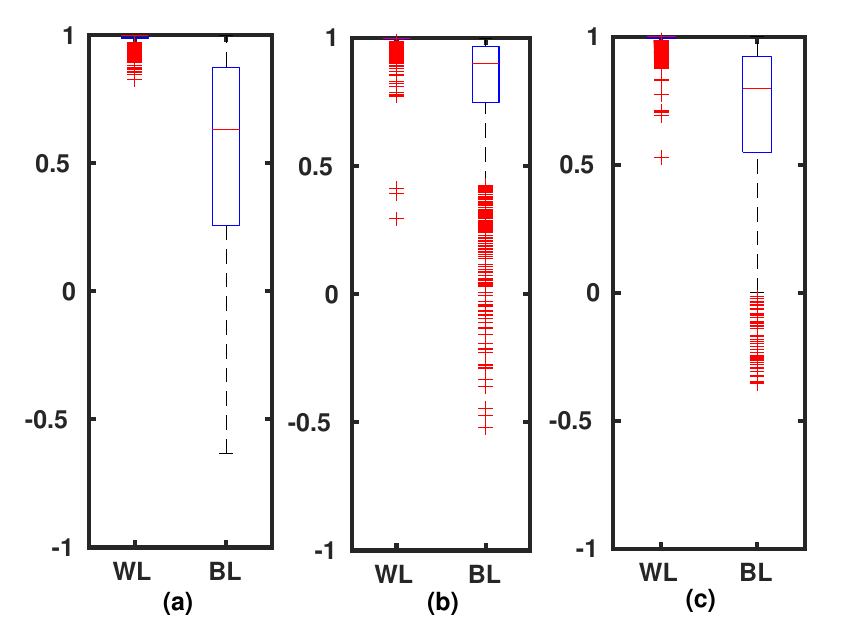}
 \caption{The Cosine score distribution between WL and BL pairs for GUE, TAE, and TEE language pairs, respectively. The obtained EERs are $7.1$, $12.1$ and $8.05$ with $N=200$, respectively.}
 \label{f12}
 \end{figure}

Initially, the x-vector frameworks for three language pair are trained using the available training partition utterances. As to train x-vector framework require voiced language-specific features, hence after MFCC feature extraction, the ground truth language labels are used to pool out language-specific features. Using the pooled feature vectors, considering $N=200$, the x-vector framework is trained for $100$ epochs. Observing the validation loss and accuracy the model belonging to the (54th, 29th, and 26th) epoch is chosen for x-vector extraction for the GUE, TAE, and TEE language pairs, respectively. After that, the discrimination ability is evaluated using the test set. It is observed that, in contrast to synthetic data, the x-vectors extracted from the $X_b$ output perform better than $X_a$, with LDA and WCCN along with cosine distance. The LDA dimension used for this study is $150$. The obtained EERs for GUE, TAE, and TEE language pairs are $7.1\%$, $12.1\%$, and $8.05\%$, respectively. The discrimination ability for all three language pairs is shown in Figure~\ref{f12}.

The LD with fixed segmentation is performed, using the test set by following the same framework used for synthetic data. Only during clustering cosine distance is used instead of GPLDA, as the discrimination is better observed with cosine distance as compared to GPLDA.

The LD with change point-inspired segmentation is performed with the test set by following a similar framework as with the synthetic dataset. The change detection and clustering are performed by using cosine distance instead of GPLDA. To optimize the hyper-parameter of the change detection framework, the change detection performance is evaluated in the first $100$ utterances in the test set. The optimal hyper-parameters ($\alpha$, $\delta$, and $\gamma$) with $N=200$ for GUE, TAE, and TEE are ($0.3$, $4.5$, and $1.1$), ($0.3$, $4.5$, and $1.1$) and ($0.3$, $3.9$, and $1.1$), respectively. With $N=50$, the hyper-parameters are ($0.3$, $0.9$, and $1.1$), ($0.3$, $0.9$, and $1.3$) and ($0.3$, $0.5$, and $1.3$), respectively. The obtained change point performances are tabulated in Table~\ref{CS-PD}.

 \begin{table}[]
\centering
\caption{Performance of the language diarization with fixed segmentation using the practical dataset, AVG: represents average across language pairs.}
\label{fs-PD}
\fontsize{7pt}{7pt}\selectfont
\begin{tabular}{|c|cccccccc|}
\hline
 & \multicolumn{8}{c|}{MSCS} \\ \hline
N & \multicolumn{4}{c|}{200} & \multicolumn{4}{c|}{50} \\ \hline
 & \multicolumn{1}{c|}{GUE} & \multicolumn{1}{c|}{TAE} & \multicolumn{1}{c|}{TEE} & \multicolumn{1}{c|}{AVG} & \multicolumn{1}{c|}{GUE} & \multicolumn{1}{c|}{TAE} & \multicolumn{1}{c|}{TEE} & AVG \\ \hline
DER & \multicolumn{1}{c|}{45.16} & \multicolumn{1}{c|}{45.91} & \multicolumn{1}{c|}{45.01} & \multicolumn{1}{c|}{\textbf{45.36}} & \multicolumn{1}{c|}{30.13} & \multicolumn{1}{c|}{31.60} & \multicolumn{1}{c|}{29.22} & \textbf{30.31} \\ \hline
JER & \multicolumn{1}{c|}{65.59} & \multicolumn{1}{c|}{66.56} & \multicolumn{1}{c|}{66.05} & \multicolumn{1}{c|}{\textbf{66.06}} & \multicolumn{1}{c|}{53.70} & \multicolumn{1}{c|}{55.74} & \multicolumn{1}{c|}{54.80} & \textbf{54.74} \\ \hline
\end{tabular}
\end{table}

\begin{table}[]
\centering
\caption{Performance of the language diarization with change point inspired segmentation using the practical dataset}
\label{CS-PD}
\fontsize{7pt}{7pt}\selectfont
\begin{tabular}{|cc|c|c|c|c|c|c|c|}
\hline
\multicolumn{2}{|c|}{Dataset}                                               & N   & IDR            & MR    & FAR   & Dm   & DER            & JER            \\ \hline
\multicolumn{1}{|c|}{\multirow{8}{*}{MSCS}} & \multirow{2}{*}{GUE} & 200 & 46.69          & 46.96 & 6.35  & 0.5  & 35.35          & 58.47          \\ \cline{3-9} 
\multicolumn{1}{|c|}{}                        &                      & 50  & 54.94          & 31.13 & 13.93 & 0.32 & 29.95 & 55.25 \\ \cline{2-9} 
\multicolumn{1}{|c|}{}                        & \multirow{2}{*}{TAE} & 200 & 50.17          & 39.86 & 9.97  & 0.5  & 36.2           & 59.74          \\ \cline{3-9} 
\multicolumn{1}{|c|}{}                        &                      & 50  & 52.46          & 20.21 & 27.33 & 0.26 & 27.75 & 53.22 \\ \cline{2-9} 
\multicolumn{1}{|c|}{}                        & \multirow{2}{*}{TEE} & 200 & 47.69          & 45.7  & 6.61  & 0.56 & 35.85          & 59.88          \\ \cline{3-9} 
\multicolumn{1}{|c|}{}                        &                      & 50  & 51.75          & 25.07 & 23.18 & 0.28 & 27.21 & 53.5  \\ \cline{2-9} 
\multicolumn{1}{|c|}{}                        & \multirow{2}{*}{AVG} & 200 & 48.18 & 44.17 & 7.65  & 0.52 & \textbf{35.8}  & \textbf{59.36} \\ \cline{3-9} 
\multicolumn{1}{|c|}{}                        &                      & 50  & 53.05& 25.47 & 21.48 & 0.28 & \textbf{28.30} & \textbf{53.99} \\ \hline
\end{tabular}
\end{table}

The performance of the LD with fixed segmentation is tabulated in Table~\ref{fs-PD}. From the table, it can be observed that the performance of LD with $N=200$ by considering the language pair GUE, TAE, and TEE are  $45.16$, $45.91$, and $45.01$ in terms of DER and $65.59$, $66.56$ and $66.05$ in terms of JER, respectively. The average performance across the language pairs is $45.36$ and $66.05$. The performance of the LD with change point-inspired segmentation is tabulated in Table~\ref{CS-PD}. From the table, it can be observed that the average performance across the language pairs with $N=200$ in terms of DER and JER is $35.8$ and $59.36$, respectively. The improvement in performance shows the significance of change point-inspired segmentation over fixed segmentation. However, the performance is far from  the performance obtained using synthetic data, i.e. $17.58$ and $29.39$ using fixed segmentation and $11.16$ and $20.61$ using change-point inspired segmentation.

\begin{figure}
 \centering
\includegraphics[height= 150pt,width= 230pt]{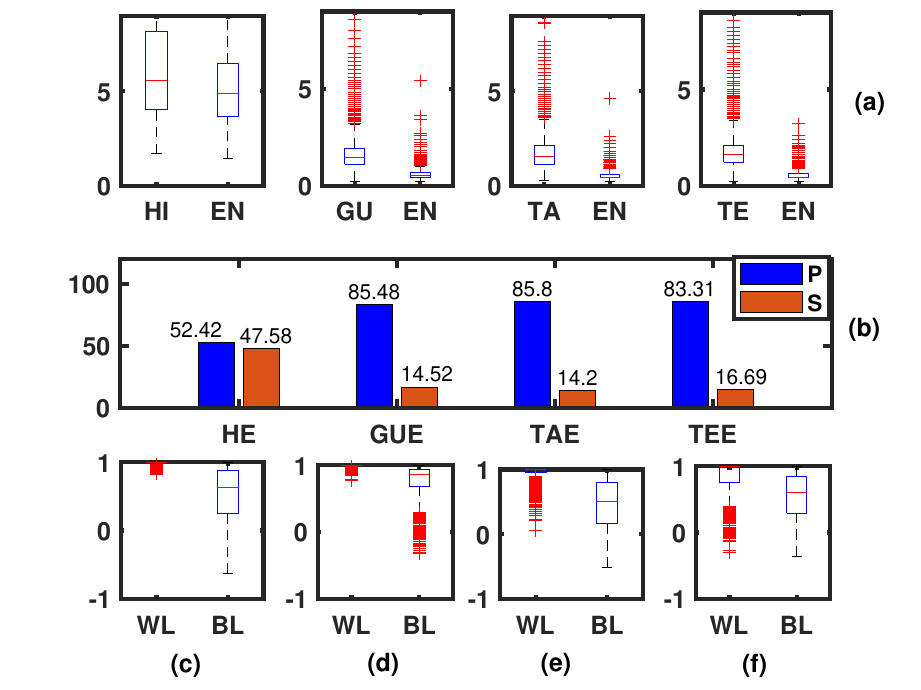}
 \caption{(a) Distribution of the monolingual segment duration of TTSF-LD (HE) and MSCS test set, (b) Percentage of primary and secondary language duration in the train set of TTSF-LD (HE) and MSCS dataset, P, and S represents primary and secondary language, (c-f) The cosine score distribution between WL and BL
pairs for GUE language pair of MSCS, at $N=200$, $150$, $100$ and $50$ with EER of (c) $7.1$, (d) $9.8$,(e) $12.8$, and (f) $29.2$, respectively.}
 \label{f13}
 \end{figure}

The performance gap is mostly due to the distributional difference in the monolingual segment duration of the secondary language between the practical and synthetic data sets. It can be observed from Figure~\ref{f13} (a), that the median of the distribution of monolingual segmentation duration of secondary language is approximately $0.5$ seconds, in contrast to approximately $3$ seconds in the case of the synthetic dataset. Hence, the use of the analysis window length of $200$ (approx. $2$  seconds duration) smooths out the contour and detects the small duration secondary language segments as the primary language and leading to the increase in the value of DER and JER. The same can be observed from Table~\ref{CS-PD}, that the MR is very high considering $N=200$. Hence even though discrimination ability reduces with a decrease in $N$, the distribution of the secondary language's segment duration encourages observing the performance by using $N=50$.

The observed LD performance with fixed segmentation case is tabulated in Table~\ref{fs-PD} and change point inspired segmentation is tabulated in Table~\ref{CS-PD}. From the table, it can be observed that the performance improves by considering $N=50$, even though the discrimination is poor. The discrimination ability by taking the GUE language pair as a case study can be observed from Figure~\ref{f13}(c-f). The obtained EERs with $N=200$, $150$, $100$ and $50$ are $7.1\%$, $9.8\%$, $12.8\%$ and $29.2\%$, respectively. The obtained average DER and JER across language pairs with $N=50$ by considering fixed segmentation are   
$30.31$ and $54.74$, in comparison with the obtained performance with $N=200$ is $45.36$ and $66.06$, respectively. Similarly, average DER and JER across language pairs with $N=50$ by considering change point-inspired segmentation are   
$28.3$ and $53.99$, in comparison with the obtained performance with $N=200$ is $35.8$ and $59.36$, respectively. This suggests the performance of the LD depends on the selection of the analysis window. Further, the selection of the analysis window not only depends on the language discrimination ability but also depends on the distribution of the monolingual segment duration. In practical utterances, by nature, the duration of the monolingual segment is small (approx. $500$ msec), and the distribution of the smaller duration segments is higher in the secondary language. Therefore, the aim should be to improve the language discrimination ability by considering a smaller analysis window duration.

\subsubsection{End-to-End Implicit LD}
The E2E LD is performed with the MSCS dataset using a similar framework used for the synthetic dataset. It is observed that the duration of the analysis window plays an important role in the LD performance, and the aim is to obtain the language discrimination in a smaller analysis window duration. Inspired by the work reported in~\cite{liu2021end} and the fact that the self-attention block of the E2E architectures captures global utterance level temporal dynamics, this work uses $N=20$ ($200$ msec) with a shift of $20$ (for capturing segment-level information) to train the E2E architecture. The architectures are trained separately using the training set of GUE, TAE, and TEE language pairs. Each architecture is trained for $100$ epochs and, the model parameters corresponding to the epoch, that provide the best validation accuracy are chosen for testing.

The performance is evaluated with four measures, i.e. along with DER and JER, identification accuracy (IDA), and equal error rate (EER). The use of IDA and EER will confirm the system's performance for the sub-utterance level language identification (SLID) task~\cite{shah2020first}. Further, the measures also enable us to compare the performance of the current work with the works reported in~\cite{liu2021end,shah2020first,rangan2020exploiting}.

\begin{table}[]
\centering
\caption{Performance of the language diarization with E2E framework using the practical dataset}
\label{E2E-PD}
\begin{tabular}{|c|cccc|}
\hline
\multicolumn{1}{|l|}{\multirow{2}{*}{}} & \multicolumn{4}{c|}{MSCS}                                                                          \\ \cline{2-5} 
\multicolumn{1}{|l|}{}                  & \multicolumn{1}{c|}{GUE}  & \multicolumn{1}{c|}{TAE}  & \multicolumn{1}{c|}{TEE}  & AVG            \\ \hline
DER                                     & \multicolumn{1}{c|}{22.9} & \multicolumn{1}{c|}{22.8} & \multicolumn{1}{c|}{21.1} & \textbf{22.26} \\ \hline
JER                                     & \multicolumn{1}{c|}{60.6} & \multicolumn{1}{c|}{60.5} & \multicolumn{1}{c|}{60.1} & \textbf{60.4}  \\ \hline
IDA                                     & \multicolumn{1}{c|}{80.9} & \multicolumn{1}{c|}{81.4} & \multicolumn{1}{c|}{81.7} & \textbf{81.3}  \\ \hline
EER                                     & \multicolumn{1}{c|}{6.3}  & \multicolumn{1}{c|}{6.9}  & \multicolumn{1}{c|}{6.0}  & \textbf{6.4}   \\ \hline
\end{tabular}
\end{table}

\begin{table}[]
\centering
\caption{Confusion matrix averaged across language pairs, P, S, and Sil represent primary, secondary, and silence respectively.}
\label{E2E-C}
\begin{tabular}{|c|c|c|c|}
\hline
\multicolumn{1}{|l|}{} & P    & S & Sil  \\ \hline
P                      & 90.6 & 0 & 9.4  \\ \hline
S                      & 65.3 & 0 & 34.7 \\ \hline
Sil                    & 16.8 & 0 & 83.1 \\ \hline
\end{tabular}
\end{table}

The performance of the LD is tabulated in Table~\ref{E2E-PD}. From the table, it can be observed that the average performance across language pairs, in terms of DER and JER is $22.6$ and $60.4$, respectively. The performance is improved in terms of DER, in comparison with the performance achieved using the change point-based approach, i.e. $28.30$ and $53.99$ in terms of DER and JER, respectively. However, the performance reported in terms of JER degrades. Further, the performance obtained in terms of IDA and EER, i.e. $81.3\%$ and $6.4\%$ is nearly the same as the performance reported in the work~\cite{liu2021end}. As discussed in our earlier work~\cite{mishra2022issues}, for a given utterance if the monolingual segment duration of one language is higher compared to the other, the obtained performance is biased towards the higher representing language, i.e. if an utterance has $80\%$ representation from one language and $20\%$ from the rest, even though the system not able to predict the rest languages and predict only the highest representative language, the DER will be $20\%$ (i.e. $100-80$). However, the system should ideally predict the segments belonging to all the representative languages. On the other hand, while computing performance JER provides equal weight to each language segment~\cite{ryant2018first,mishra2022importance}. Hence, if some language segments are not predicted by the system, the JER is higher. The same can be observed here from the confusion matrix tabulated in Table~\ref{E2E-C}. The confusion matrix shows that the system is predicting only the primary language and is not able to predict the secondary language. Additionally, by nature, the CS utterances have the primary language segment duration much higher than the secondary (observed from Figure~\ref{f13}(a)). Therefore, in the obtained performance, though the DER decreases the JER increases.

The degradation in performance in terms of JER is primarily due to acoustic similarity (in $200$ msec, i.e. in syllable/word level mainly the secondary language is produced by using the primary language phoneme sequence) and imbalance in the training data (observed from Figure~\ref{f13}(b)). The network uses a discriminative strategy for training, and due to the scarcity of sufficient training data from secondary language and acoustic similarity, the system performance is biased towards the primary language. However, the same framework works well with the synthetic dataset. This is due to two reasons, (1) considering the analysis window length $N=200$ (approx. $2$ seconds) minimizes the effect of acoustic similarity, and forces the network to learn the language discrimination, (2) equally distributed training data (observed from Figure~\ref{f13}(b)). The nature of practical CS utterances (1) imbalance training data (2) the small duration of the secondary language's segment duration poses a challenge for the training of the E2E framework to perform LD. The issue can be resolved by minimizing the effect of data imbalance and enabling the framework to extract the language discriminative evidence with a small analysis window length.

\section{Self-supervised representation for implicit language diarization}
\label{SSL}
The W2V framework pre-trained with $23$ Indian languages, with considering $10,000$ hours of unlabeled speech data is considered here. The pre-trained framework is fine-tuned separately, with the training data available for each language pair in the MSCS dataset. The fine-tuned framework is used as a feature extractor to obtain the language representations. Further, the language representations are used to perform LD with fixed segmentation, change-point-based segmentation, and E2E frameworks. The detailed descriptions are given in the following subsections.
\subsection{W2V Pre-training}
The pre-training architecture used in this study is originally proposed in~\cite{gupta2021clsril}, to perform automatic speech recognition tasks in Indian languages. The training of pre-training architecture can be divided into four stages, (1) obtaining latent representation from the speech signal using CNN, (2) quantizing the latent representations, (3) prediction of masked representations using self-attention transformers, (4) computation of contrastive and divergence loss. 

 \begin{figure}
 \centering
\includegraphics[height= 150pt,width= 230pt]{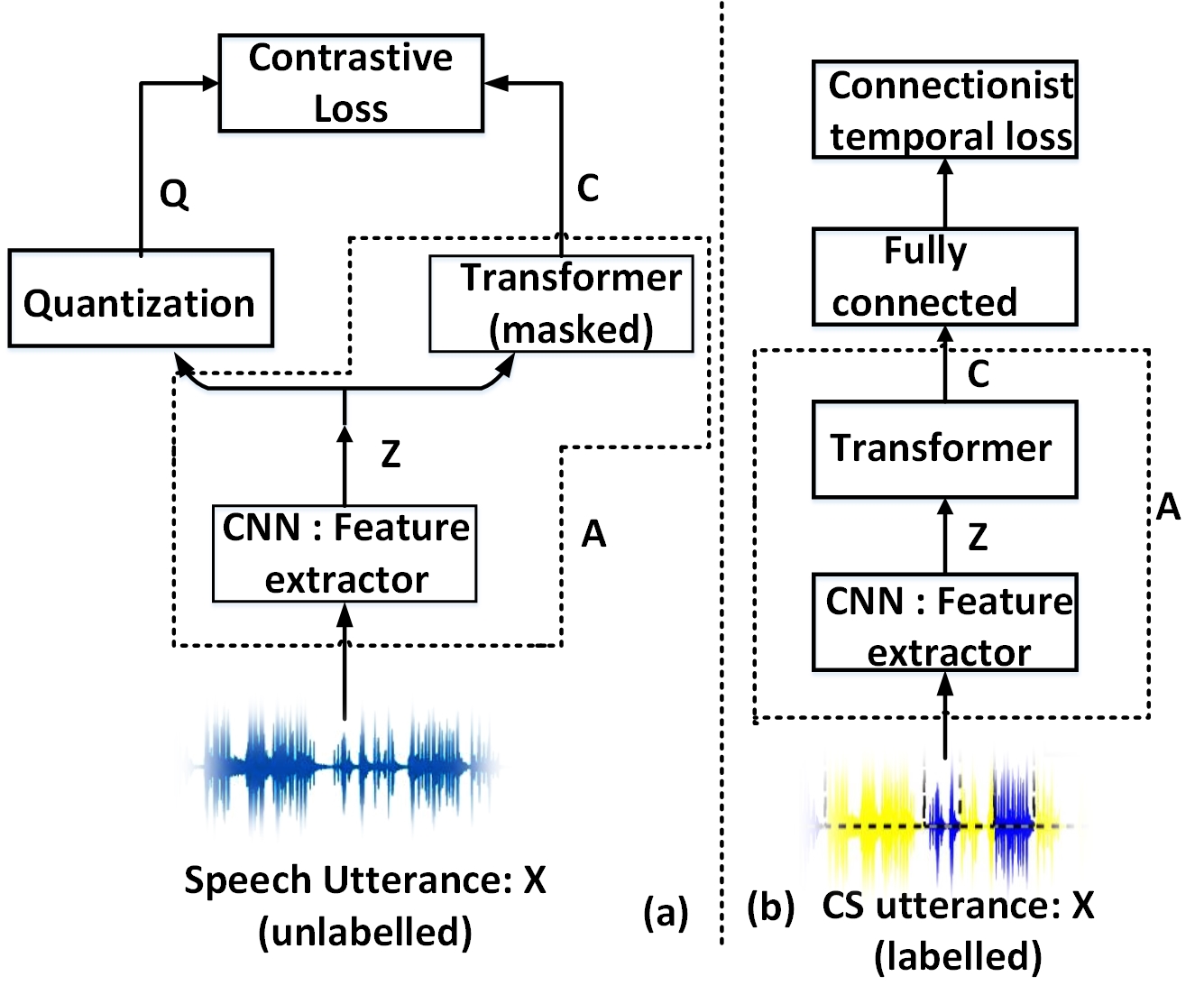}
 \caption{Wav2vec (a) pre-training and (b) fine-tuning architecture.}
 \label{f14}
 \end{figure}

The architecture has $7$ $1D$-CNN layers having $512$ filters each, kernel size of ($10,3,3,3,3,2,2$) and stride ($5,2,2,2,2,2,2$), followed by $12$ self-attention transformer layers. Initially, zero mean amplitude normalized speech signal ($X$) is passed through the CNN layers to obtain the latent representation ($Z$). The latent representations are computed in every $20$ msec with a receptive field of $25$ msec. After that, the latent representations are quantized using $G$ codebooks, each having $V$ number of quantized vectors of dimension $d/G$. Given a latent representation $z \in Z$, the Gumbel softmax is used to decide on the nearest quantization vector in each codebook and then concatenated to form a $d$ dimensional quantized vector ($q \in Q$). On the other side, the latent representations passed through the self-attention-based transformer encoders with masking (having a minimum $M=16$ consecutive frames) to obtain the contextualized representation ($C$). Suppose the masked contextualized representation and corresponding quantized representation are represented as $c_{m}$ and $q_{m}$, respectively. Further, the $k$-distractors' contextualized representations and quantized representations randomly sampled from the other segments of the utterance are represented as $c'$ and $q'$. The architecture is trained using contrastive and divergence loss ($L_c+\alpha L_d$) mentioned in Equation~\ref{cl} and \ref{div}, where $\varphi(.)$ represents the cosine similarity. The contrastive loss ensures the correct prediction of masked contextualized representations, by increasing the similarity between the $c_{m}$ and $q_{m}$ and increasing the divergence between $C'$ and $Q'$. The divergence loss (i.e. entropy/uncertainty) ensures that the quantized vectors should not be biased towards any one codebook/ quantized vector. The summary of pre-training is depicted in Figure~\ref{f14}. The detailed procedure of training can be found at ~\cite{gupta2021clsril} and ~\cite{baevski2020wav2vec}. 

\begin{equation}
\label{cl}
L_{c} =  - log    \dfrac{e^{\varphi(c_{m}, q_{m})/k}}{\sum_{q'\in Q' }  e^{\varphi(c',q')/k}} 
\end{equation}

\begin{equation}
    \label{div}
    L_{d}=\frac{1}{GV}\sum_{g=1}^{G}\sum_{v=1}^{V}\Bar{p}_{g,v}\log(\bar{p}_{g,v})
\end{equation}
The pre-training is performed with $23$ Indian languages considering the minimum masking duration of $320$ msec ($16*20$). Hence, during training irrespective of the trained languages, the architecture is tuned up by learning the temporal dynamics between the acoustic units to predict the syllable/word level representation.

\subsection{W2V Fine-tuning}

After pre-training, as shown in Figure~\ref{f14} the CNN and transformer layers (marked as "A " block in Figure~\ref{f14}(a-b)) are detached from the pre-training architecture, and a fully connected layer $5$ ($2$ language, $1$ for silence and rest $2$ is for blank and hyphen) neurons with softmax activation are added to it. The fine-tuning architecture is shown in Figure~\ref{f14}(b).

The pre-trained architecture is expected to capture the long-term temporal dynamics and be capable of predicting language-independent syllables/words. Hence, it is expected during fine-tuning, the pre-training architecture will guide the fine-tuned architecture to discriminate between the languages. Further, as the segmental language labels are available, to capture better temporal context for language discrimination, connectionist temporal classification (CTC) loss is used. For fine-tuning the CTC loss is computed by comparing the available language tags at each $200$ msec with the language tag predicted by the softmax at each $20$ msec. For each language pair, the architectures are finetuned for $900$ epochs. With respect to the validation loss, the optimal models are chosen for each language pair to extract the implicit W2V features.

\subsection{W2V Representation}
After fine-tuning, the fully connected layer is detached from the architecture and the "A" block consists of CNN, and transformer layers are used as a feature extractor. In contrast to MFCC features computed in each $10$ msec, the W2V features are computed in each $20$ msec. Motivating by the frame aggregation study of explicit language representation, the W2V features are statistically pooled with a given analysis window length $N$ to obtain the language representations. As the W2V features are extracted in each $20$ msec, the analysis window length ($N_{w}$) used to extract language representation is half the analysis window length followed for the x-vector representations (i.e. $N_{w}=\frac{N}{2}$). Hence, for easy readability and comparison, even though the frame aggregation is performed with $N_{w}$, it will be represented throughout the work in terms of $N$ (i.e. $2 \times N_{w}$).

 \begin{figure*}
 \centering
\includegraphics[height= 130pt,width= 500pt]{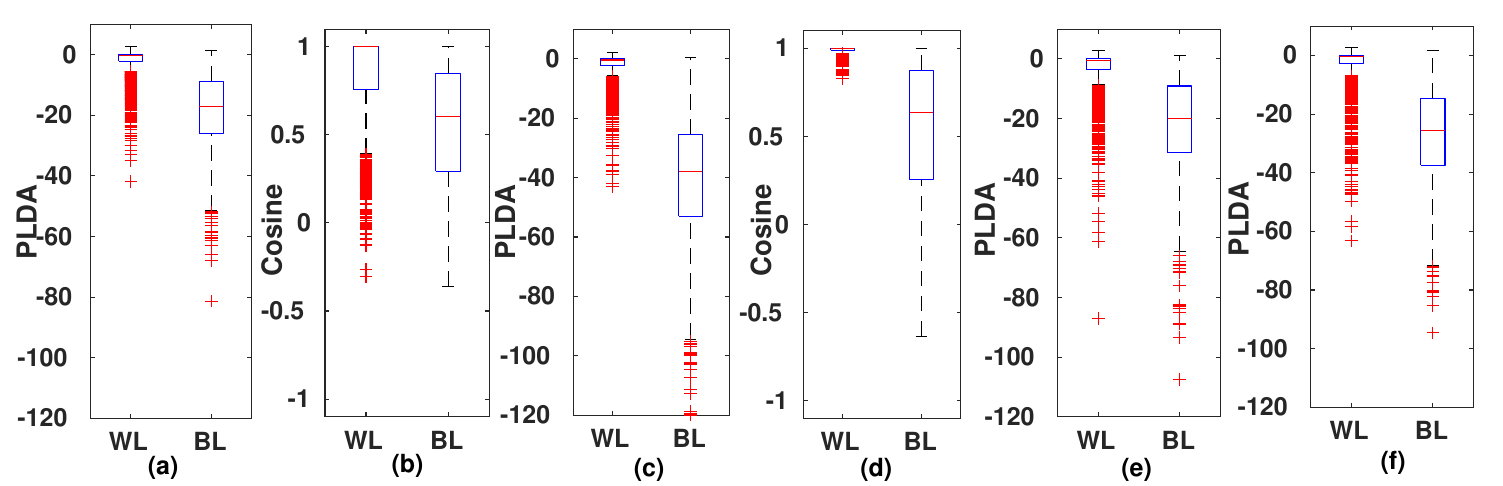}
 \caption{WL and BL score distribution (a) W2V, (b) x-vector representation with $N=50$, (c) W2V, (d) x-vector representation with $N=200$ for GUE language pair, (e) W2V for TAE language pair, (f) for TEE language pair with $N=50$. The obtained EERs are $14\%$, $29.2\%$, $5.05\%$, $7.1\%$, $17.75\%$ and $13.35\%$, respectively.  }
 \label{f15}
 \end{figure*}

The GUE language pair is used to observe the language discrimination ability of the W2V language representation. The study is performed by using, with LDA and WCCN/without LDA and WCCN, and considering cosine distance/ PLDA distance. The study shows that the performance is better by considering LDA and WCCN of LDA dimension $25$, along with PLDA as a distance matrix. The PLDA score distributions of the WL and BL pairs for $N=50$ and $N=200$ are shown in Figure~\ref{f15}(a) and (c). For comparison, the cosine score distributions using the x-vector as the representation are depicted in Figure~\ref{f15}(b) and (d). From the figure, it is observed that the W2V representation provides better discrimination as compared to the x-vector. The same is also evident by evaluating EER as the objective measure. Using W2V as language representation the obtained EERs with $N=50$ and $200$ are $14\%$ and $5.05\%$, in contrast to $29.2\%$ and $7.1\%$ using x-vector as representation, respectively. Though the discrimination ability is better with considering $N=200$, due to the distribution of monolingual segment duration (i.e. median of $500$ msec) of secondary language in the MSCS dataset, $N=50$ is preferable over $N=200$ to avoid segment smoothing for performing LD. With $N=50$ the jump in language discrimination performance from $29.2$ to $14$ may help in improving LD performance. With $N=50$, using W2V representation the PLDA score distributions of the WL and BL pairs of TAE and TEE language pair are depicted in Figure~\ref{f15}(e) and (f). The obtained EERs for the TAE and TEE language pair are $17.75\%$ and $13.35\%$, respectively. 

\subsection{Fixed and change point inspired segmentation}

The LD with fixed segmentation is performed using W2V as a representation vector.
For a given test utterance, the W2V features are extracted from the fine-tuned W2V model. After extracting the features, the frame aggregation is performed (statistical pooling) with $N=50$ and a shift of one feature vector, then the aggregated vectors are projected to $25$ dimensional projected space using the trained LDA and WCCN matrix. The projected vectors are clustered using AHC with GPLDA as a distance matrix.  The obtained performance in terms of DER and JER is tabulated in Table~\ref{w2v_FS}. From the table, it is observed that the average performance across language pairs in terms of  DER and JER are $18.74\%$ and $33.24\%$, respectively. The obtained performance is higher than the best performance achieved with the x-vector representation (i.e. $22.26\%$ and $60.4\%$ in terms of DER and JER using the E2E framework). This shows the significance of W2V language representation over the x-vector representation.

\begin{table}[]
\centering
\caption{Performance of LD with fixed segmentation framework using W2V representation.}
\label{w2v_FS}
\begin{tabular}{|c|cccc|}
\hline
    & \multicolumn{4}{c|}{MSCS}                                                                             \\ \hline
N   & \multicolumn{4}{c|}{50}                                                                               \\ \hline
    & \multicolumn{1}{c|}{GUE}   & \multicolumn{1}{c|}{TAE}   & \multicolumn{1}{c|}{TEE}   & AVG            \\ \hline
DER & \multicolumn{1}{c|}{19.08} & \multicolumn{1}{c|}{19.37} & \multicolumn{1}{c|}{17.78} & \textbf{18.74} \\ \hline
JER & \multicolumn{1}{c|}{34.12} & \multicolumn{1}{c|}{33.44} & \multicolumn{1}{c|}{32.17} & \textbf{33.24} \\ \hline
\end{tabular}
\end{table}

The change point-based LD is performed using W2V representation. The same framework used with x-vector-based representation is also used here. Initially, the hyperparameters of the change detection framework are optimized by evaluating the performance of change detection on the first $100$ test trails. The optimal hyper-parameters ($\alpha$, $\delta$, and $\gamma$) to perform the change detection task for GUE, TAE, and TEE language pairs are ($0.9$, $0.9$, and $0.5$), ($0.7$, $0.9$, and $0.7$) and ($0.9$, $0.9$, and $0.7$), respectively. Using the change point information the clustering is performed and then the predicted RTTMs are generated. The obtained LD performance along with the change detection performance is tabulated in Table~\ref{w2v_CP}.

\begin{table}[]
\centering
\caption{Performance of LD with change point based segmentation framework using W2V representation.}
\label{w2v_CP}
\fontsize{7pt}{7pt}\selectfont
\begin{tabular}{|c|c|c|c|c|c|c|c|c|}
\hline
                      &     & N  & IDR   & MR    & FAR   & Dm   & DER            & JER            \\ \hline
\multirow{4}{*}{MSCS} & GUE & 50 & 70.6  & 17.69 & 11.71 & 0.13 & 11.82          & 30.00          \\ \cline{2-9} 
                      & TAE & 50 & 71.05 & 11.7  & 17.25 & 0.13 & 10.93          & 28.22          \\ \cline{2-9} 
                      & TEE & 50 & 71.47 & 16.05 & 12.48 & 0.12 & 10.49          & 28.26          \\ \cline{2-9} 
                      & AVG & 50 & 71.04 & 15.14 & 13.81 & 0.13 & \textbf{11.08} & \textbf{28.82} \\ \hline
\end{tabular}
\end{table}

The obtained average performance across the language pairs for the change detection task in terms of IDR, MR, FAR, and $D_{m}$ is $71.04\%$, $15.14\%$, $13.81\%$ and $0.13$ seconds, respectively. In comparison with the x-vector representation, the IDR improves from $53.05\%$ to $71.04\%$. The effect of the same is also observed in the performance of LD. The obtained average LD performance in terms of DER and JER is $11.08$ and $28.82$, respectively. The obtained performance is better as compared to the performance obtained using fixed segmentation with W2V representation (i.e. $18.74$ and $33.24$ in terms of DER and JER) and the best-performing x-vector-based E2E framework (i.e. $22.26$ and $60.4$ in terms of DER and JER using the E2E framework). Motivated by the improvement of the performance by using W2V representation with fixed and change point-based framework, the W2V representations are used with the E2E framework, with an expectation to further improve the performance.

\subsection{E2E diarization}

 \begin{figure}
 \centering
\includegraphics[height= 190pt,width= 200pt]{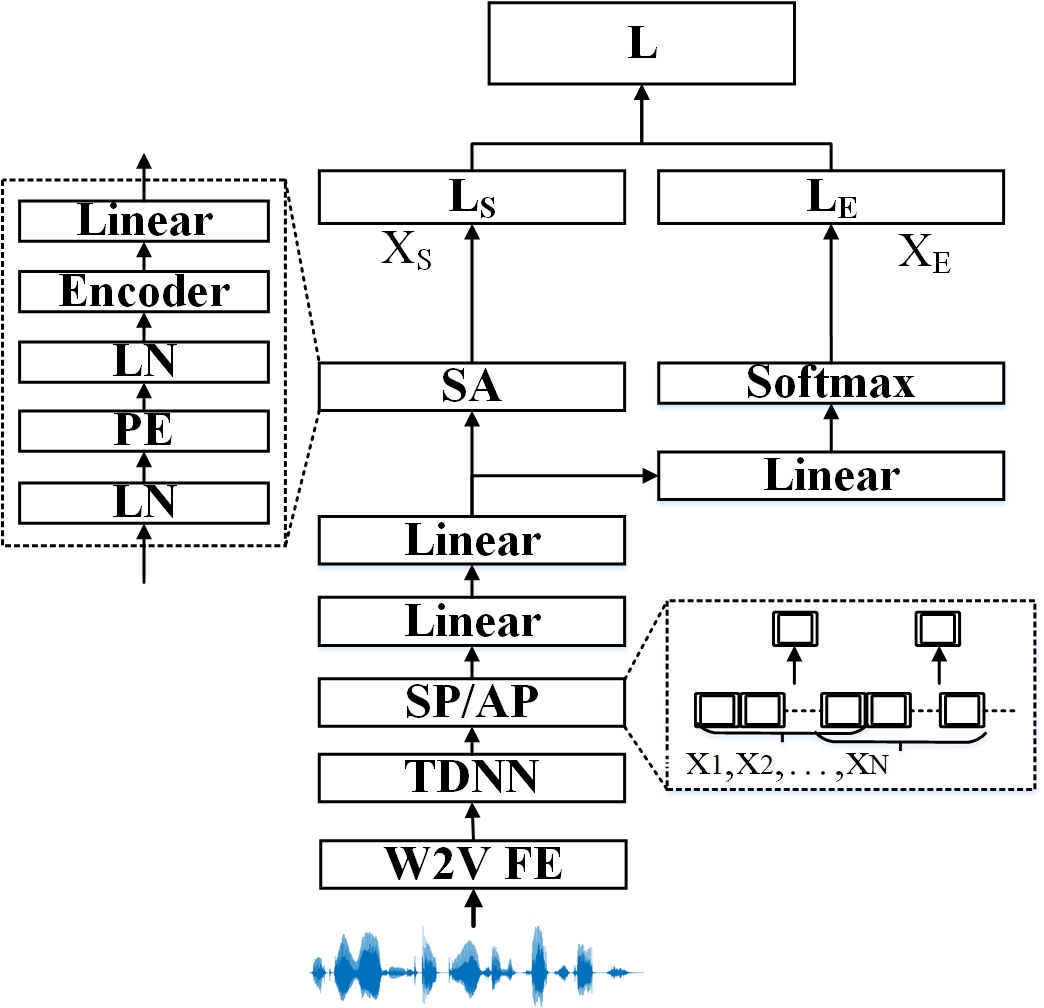}
 \caption{W2V based E2E diarization framework, SP: statistical pooling, AP: attention pooling, SA: self-attention, LN: length normalization, and PE: positional encoding}
 \label{f16}
 \end{figure}

The E2E framework used with x-vector representation to perform LD using the MSCS dataset is used here with a small modification. Instead of MFCC features, the features extracted from the W2V are given input to the TDNN layer. The same is depicted in Figure~\ref{f16}. The W2V FE is not further trained (detached from the gradient update) during E2E training. As the features are extracted with each $20$ msec, the $N_{w}$ is considered here as $10$ (i.e. same as $N=20$). For each language pair, three E2E frameworks are trained (1) considering the pre-trained W2V network as a feature extractor with frame aggregation through statistical pooling (P-W2V-SP), (2) considering the fine-tuned network as a feature extractor with frame aggregation through statistical pooling (F-W2V-SP), (3) considering the fine-tuned network as a feature extractor with frame aggregation through attention pooling (F-W2V-AP). It is observed in the literature that frame-aggregation through attention pooling (AP) captures better sequence representation in comparison to statistical pooling (AP)~\cite{bai2021speaker}, hence motivated to use it here. For each language pair, each network is trained for $100$ epochs. The models that belong to the epoch providing the least validation loss are used for testing. Further, the performance is also evaluated, directly from the W2V finetuned architecture (W2V-F). For the given test CS utterance, it is assumed that each label, in the predicted label sequence (by the CTC decoder) is  predicted in each $200$ msec for generating the predicted RTTM file.

The evaluated performances using test set CS utterances for each language pair are tabulated in Table~\ref{e2e_w2v_per}. The average performance across language pairs with the W2V-F framework in terms of DER and JER is $24.3$ and $36.8$, respectively. Using P-W2V-SP, the average performance is $20.3$ and $47.2$, respectively. Though the DER decreases, the JER increases shows that the language-specific finetuning helps in improving performance. The performance of F-W2V-SP in terms of DER and JER is $11.9$ and $22.3$, respectively. This shows the significance of using the E2E framework with language-specific finetuned implicit representation. Further, the performance is improved by using AP instead of SP. The obtained performance using F-W2V-AP in terms of DER and JER is $11.2$ and $21.8$, respectively. The performance of LD using the E2E framework is comparable with the performance obtained for the change point segmentation-based framework in terms of DER (i.e. $11.08$). However, in terms of JER the performance of the E2E framework is better than the change point segmentation-based framework (i.e. $28.82$).

\begin{figure}
 \centering
\includegraphics[height= 100pt,width= 240pt]{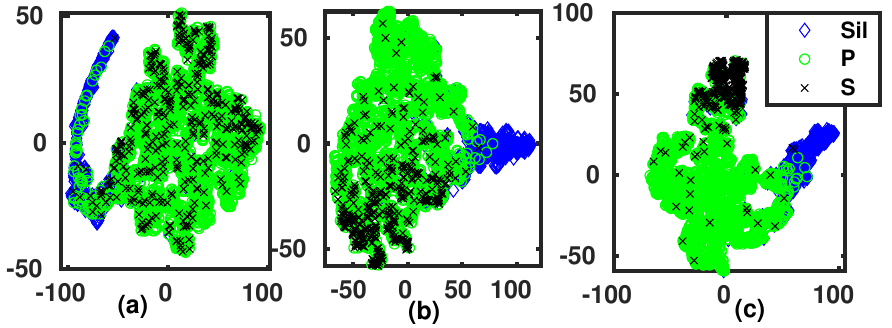}
 \caption{The t-SNE distribution of embeddings using TAE language pair obtained from  (a) X-E2E, (b) P-W2V-SP, and (d) F-W2V-AP architectures, respectively.}
 \label{f17}
 \end{figure}

\begin{table*}[]
\centering
\caption{Performance of E2E LD using W2V representation on MSCS dataset, W2V-F: fine-tuned W2V model, P-W2V-SP: E2E framework using pretraining W2V feature with statistical pooling, F-W2V-SP and F-W2V-AP: E2E framework using fine-tuned W2V feature with statistical and attention pooling, respectively.}
\label{e2e_w2v_per}
\begin{tabular}{|c|c|cccc|cccc|cccc|cccc|}
\hline
\multicolumn{1}{|l|}{\multirow{2}{*}{}} &
  \multicolumn{1}{l|}{\multirow{2}{*}{}} &
  \multicolumn{4}{c|}{W2V-F} &
  \multicolumn{4}{c|}{P-W2V-SP} &
  \multicolumn{4}{c|}{F-W2V-SP} &
  \multicolumn{4}{c|}{F-W2V-AP} \\ \cline{3-18} 
\multicolumn{1}{|l|}{} &
  \multicolumn{1}{l|}{} &
  \multicolumn{1}{c|}{IDR} &
  \multicolumn{1}{c|}{EER} &
  \multicolumn{1}{c|}{DER} &
  JER &
  \multicolumn{1}{c|}{IDR} &
  \multicolumn{1}{c|}{EER} &
  \multicolumn{1}{c|}{DER} &
  JER &
  \multicolumn{1}{c|}{IDR} &
  \multicolumn{1}{c|}{EER} &
  \multicolumn{1}{c|}{DER} &
  JER &
  \multicolumn{1}{c|}{IDR} &
  \multicolumn{1}{c|}{EER} &
  \multicolumn{1}{c|}{DER} &
  JER \\ \hline
\multirow{4}{*}{MSCS} &
  GUE &
  \multicolumn{1}{c|}{82.2} &
  \multicolumn{1}{c|}{5.3} &
  \multicolumn{1}{c|}{23.7} &
  35.4 &
  \multicolumn{1}{c|}{83.4} &
  \multicolumn{1}{c|}{5.5} &
  \multicolumn{1}{c|}{20.2} &
  44.0&
  \multicolumn{1}{c|}{90.0} &
  \multicolumn{1}{c|}{3.3} &
  \multicolumn{1}{c|}{12.1} &
  22.8 &
  \multicolumn{1}{c|}{90.0} &
  \multicolumn{1}{c|}{3.3} &
  \multicolumn{1}{c|}{11.3} &
  22.4 \\ \cline{2-18} 
 &
  TAE &
  \multicolumn{1}{c|}{80.9} &
  \multicolumn{1}{c|}{5.6} &
  \multicolumn{1}{c|}{25.0} &
  37.2 &
  \multicolumn{1}{c|}{82.4} &
  \multicolumn{1}{c|}{5.8} &
  \multicolumn{1}{c|}{21.6} &
  50.8 &
  \multicolumn{1}{c|}{89.4} &
  \multicolumn{1}{c|}{3.5} &
  \multicolumn{1}{c|}{13.0} &
  23.8 &
  \multicolumn{1}{c|}{89.8} &
  \multicolumn{1}{c|}{3.5} &
  \multicolumn{1}{c|}{11.8} &
  23.2 \\ \cline{2-18} 
 &
  TEE &
  \multicolumn{1}{c|}{82.9} &
  \multicolumn{1}{c|}{5.1} &
  \multicolumn{1}{c|}{24.2} &
  37.8 &
  \multicolumn{1}{c|}{83.6} &
  \multicolumn{1}{c|}{5.4} &
  \multicolumn{1}{c|}{19.2} &
  47.0 &
  \multicolumn{1}{c|}{90.4} &
  \multicolumn{1}{c|}{3.1} &
  \multicolumn{1}{c|}{10.7} &
  20.5 &
  \multicolumn{1}{c|}{90.5} &
  \multicolumn{1}{c|}{3.1} &
  \multicolumn{1}{c|}{10.7} &
  20.0 \\ \cline{2-18} 
 &
  AVG &
  \multicolumn{1}{c|}{82} &
  \multicolumn{1}{c|}{5.3} &
  \multicolumn{1}{c|}{\textbf{24.3}} &
  \textbf{36.8} &
  \multicolumn{1}{c|}{83.1} &
  \multicolumn{1}{c|}{5.5} &
  \multicolumn{1}{c|}{\textbf{20.3}} &
  \textbf{47.2} &
  \multicolumn{1}{c|}{89.9} &
  \multicolumn{1}{c|}{3.3} &
  \multicolumn{1}{c|}{\textbf{11.9}} &
  \textbf{22.3} &
  \multicolumn{1}{c|}{90.1} &
  \multicolumn{1}{c|}{3.3} &
  \multicolumn{1}{c|}{\textbf{11.2}} &
  \textbf{21.8} \\ \hline
\end{tabular}
\end{table*}
The obtained confusion matrices from the E2E frameworks are tabulated in Table~\ref{conf-fi}. It is observed from the confusion matrix that the X-E2E framework provides biased performance toward the primary language, as it cannot predict the secondary language. Using pre-trained W2V representation in P-W2V-SP, the primary language bias is somewhat reduced. The framework predicts $28.8\%$ secondary language segment as a secondary language and $45.69\%$ of the secondary language segment as the primary language. This shows the significance of a pre-trained W2V network as a non-linear feature extractor, that captures temporal dynamics of speech signal by predicting the syllable/word level representation. Hence instead of using the MFCC as a feature extractor (that is only able to capture frame-level information), the use of pre-trained W2V features reduces the primary language bias and guides the E2E framework to discriminate between languages. The same also can be observed from the t-SNE visualization depicted in Figure~\ref{f17}(a) and (b). In both cases, though the silence can able to mostly form a separate cluster, the embeddings belonging to primary and secondary languages are overlapped. As the primary language bias reduces from the X-E2E to P-W2V-SP framework, the overlap between the embeddings is also reduced.  Further, to capture the language-specific temporal dynamics, the W2V pre-trained framework is finetuned with the segmental language sequences and used as an implicit language-specific feature extractor. The extracted features are then trained with the E2E framework considering AP as a frame aggregation strategy to further reduce the primary language bias. The framework predicts $75.5\%$ secondary language segment as a secondary language and $21.6\%$ of the secondary language segment as the primary language. A similar conclusion can also be observed from the t-SNE distribution depicted in Figure~\ref{f17}(c). As compared to P-W2V-SP, the embeddings obtained from the F-W2V-AP are showing less overlapped between primary and secondary languages. This justifies the significance of using implicit W2V fine-tuned representation with the E2E framework by considering AP as a frame aggregator.

\begin{table}[hbt!]
\centering
\caption{Performance comparison using confusion matrix (averaged across language pairs), P: primary, S: secondary, and Sil: silence.}
\label{conf-fi}
\begin{tabular}{|l|l|l|l|l|}
\hline
Model                     &     & P     & S              & Sil   \\ \hline
\multirow{3}{*}{X-E2E} & P   & 90.6 & 0              & 9.3  \\ \cline{2-5} 
                          & S   & 65.2 & \textbf{0}     & 34.7 \\ \cline{2-5} 
                          & Sil & 16.8 & 0              & 83.1 \\ \hline

\multirow{3}{*}{P-W2V-SP} & P   & 89.18 & 2.8              & 8.02  \\ \cline{2-5} 
                          & S   & 45.69 & \textbf{28.8}     & 25.3 \\ \cline{2-5} 
                          & Sil & 13.09 & 1.94              & 84.97 \\ \hline

\multirow{3}{*}{F-W2V-AP} & P   & 95.3 & 3.0           & 1.6  \\ \cline{2-5} 
                          & S   & 21.6 & \textbf{75.5} & 2.8 \\ \cline{2-5} 
                          & Sil & 15.3  & 3.2           & 81.3 \\ \hline
\end{tabular}
\end{table}

\section{Discussion}
\label{dis}
Initially, with the aim to develop implicit LD that can be easily generalized for low/zero resource language, the implicit language representations are explored. It is observed from our earlier study reported in~\cite{mishra2023spoken}, the x-vector is a better implicit language representation in the CS scenario to start with. Further, In the literature on SD (a task similar to LD), the exploration is performed using three broad categories of frameworks. i.e. (1) change detection-based segmentation using acoustic features, followed by clustering, (2) fixed segmentation followed by clustering using speaker representations (i/d/x-vectors), and (3) E2E framework by joint loss of clustering, representation extraction, and VAD. Motivated by the same, this work used the x-vector as implicit language representation and performed diarization in three frameworks (1) fixed segmentation followed by clustering, (2) change-point-based segmentation followed by clustering, and (3) E2E framework. Initially, the frameworks are developed using a synthetically generated CS HE dataset (TTSF-LD).

The obtained performance with each framework is tabulated in Table~\ref{imp_exp_sum}. The performance of LD with implicit representation using the fixed segmentation framework is $17.58$ and $29.39$ in terms of DER and JER, respectively. Using change point-based segmentation is $11.16$ and $20.61$, and using the E2E framework is $5.81$ and $6.38$, respectively. This suggests the performance is improved by incorporating change point information into the LD framework and further improved using the E2E framework, by jointly optimizing language representation extraction, VAD, and self-attention(can be viewed as clustering~\cite{fujita2019end}) loss. To compare the performance of implicit LD with explicit, the performance is also evaluated using grapheme posterior-based explicit representation. The performance of the explicit LD is tabulated in Table~\ref{imp_exp_sum}. The performance of the explicit LD using the fixed segmentation framework is $12.37$ and $20.61$, respectively. Using change point-based segmentation the performance is $9.3$ and $17.23$, and using the E2E framework the performance is $11.37$ and $10.7$, respectively. This observation shows the performance of implicit LD is at par with the performance of explicit LD. However, it is observed that compared to the implicit representation, the explicit representation is able to give better language discrimination with a smaller value of $N$.

\begin{table}[]
\centering
\caption{Performance comparison of LD using synthetically generated TTSF-LD dataset with considering x-vector as language representation, FS: fixed segmentation, CPS: change point based segmentation.}
\label{imp_exp_sum}
\begin{tabular}{|c|ccc|ccc|}
\hline
\multicolumn{1}{|l|}{} & \multicolumn{3}{c|}{Implicit} & \multicolumn{3}{c|}{Explicit} \\ \hline
N                      & \multicolumn{3}{c|}{200}      & \multicolumn{3}{c|}{100}      \\ \hline
\multicolumn{1}{|l|}{} &
  \multicolumn{1}{c|}{FS} &
  \multicolumn{1}{c|}{CPS} &
  E2E &
  \multicolumn{1}{c|}{FS} &
  \multicolumn{1}{c|}{CPS} &
  E2E \\ \hline
DER &
  \multicolumn{1}{c|}{17.58} &
  \multicolumn{1}{c|}{11.16} &
  \textbf{5.81} &
  \multicolumn{1}{c|}{12.37} &
  \multicolumn{1}{c|}{9.3} &
  \textbf{11.37} \\ \hline
JER &
  \multicolumn{1}{c|}{29.39} &
  \multicolumn{1}{c|}{20.61} &
  \textbf{6.38} &
  \multicolumn{1}{c|}{20.74} &
  \multicolumn{1}{c|}{17.23} &
  \textbf{10.7} \\ \hline
\end{tabular}
\end{table}

The performance is further evaluated with a practical MSCS dataset using the proposed implicit frameworks. The averaged performance across language pairs is tabulated in Table~\ref{PD_sum}. The performances in terms of JER using the x-vector as the implicit language representation with fixed segmentation, change point segmentation, and E2E framework are $54.74$, $53.9$, and $60.4$, respectively. The performance difference between the TTSF-LD and MSCS datasets is mostly due to two reasons: (a) the imbalance in training data from primary to secondary (approximately $4:1$),  and (b) the secondary languages' monolingual segment duration. As a single speaker speaking multiple languages, mostly he adapts his primary language production system (articulator dynamics and phonemes) to produce secondary languages. In such a scenario, the required $N$ to discriminate between languages is higher. Further, due to the smaller monolingual segment duration, the use of a higher $N$ smoothed out the prediction and increased the JER. On the other hand, though the use of smaller $N$ able decreases the JER, it doesn't have the required language discrimination ability. Hence the aim is to get better language discrimination in a smaller analysis window length $N$.

\begin{table}[]
\centering
\caption{Performance comparison of Implicit LD using MSCS dataset, A: x-vector representation, B: W2V representation.}
\label{PD_sum}
\begin{tabular}{|c|cc|cc|cc|}
\hline
\multirow{2}{*}{} & \multicolumn{2}{c|}{FS}    & \multicolumn{2}{c|}{CPS}   & \multicolumn{2}{c|}{E2E}   \\ \cline{2-7} 
                  & \multicolumn{1}{c|}{A} & B & \multicolumn{1}{c|}{A} & B & \multicolumn{1}{c|}{A} & B \\ \hline
DER & \multicolumn{1}{c|}{30.31} & \textbf{18.74} & \multicolumn{1}{c|}{28.3} & \textbf{11.08} & \multicolumn{1}{c|}{22.26} & \textbf{11.2} \\ \hline
JER & \multicolumn{1}{c|}{54.74} & \textbf{33.24} & \multicolumn{1}{c|}{53.9} & \textbf{28.82} & \multicolumn{1}{c|}{60.4}  & \textbf{21.8} \\ \hline
\end{tabular}
\end{table}

A self-supervised W2V-based framework is considered, to achieve better language discrimination with a smaller value of $N$. Using self-supervision, the network is pre-trained with $23$ Indian languages, considering minimum masking length $M=16$ (approx. $320$ msec) captures the syllable/word level temporal dynamics. Further during fine-tuning with language-specific data, the network learns the syllable/word level temporal dynamics to discriminate between languages. Like explicit grapheme posterior representation, the use of the implicit way of fine-tuning captures language-specific evidence implicitly and hypothesizes to improve the language discrimination with a smaller value of $N$. The same is being observed from the language discrimination analysis and further, the use of W2V implicit representations improves the LD performance in all three frameworks. The obtained performance is tabulated in Table~\ref{PD_sum}. The obtained performances in terms of JER using fixed segmentation, change point segmentation, and E2E framework are $33.24$, $28.82$, and $21.8$, respectively.



\section{Conclusion}
\label{con}
In this study, the implicit approach is explored to perform the LD task. The performances of LD on synthetic data with the x-vector as implicit language representation using fixed segmentation, change point-based segmentation, and E2E approach is comparable with the performance achieved using explicit representation. Extending to MSCS practical dataset, it is observed that the model output is biased toward the primary language. This is due to the unavailability of sufficient secondary language training data, and secondary languages' monolingual segment duration to learn and detect the discrimination between primary and secondary. The issue is resolved to some extent by considering implicit W2V fine-tuned representations.

In the future, the framework can be further explored to achieve better discrimination between the languages. Like W2V, other domain adaptation-based frameworks can be studied to improve language representations. Further, the language representation can be improved by using a framework like GAN/VAE that can learn generative regularized language space. The regularized generative space of language representation may help in dealing with unbalanced training data and may provide better language discrimination. 
\bibliographystyle{IEEEtran}
\bibliography{sampbib}

\begin{thebibliography}{10}
\providecommand{\url}[1]{#1}
\csname url@samestyle\endcsname
\providecommand{\newblock}{\relax}
\providecommand{\bibinfo}[2]{#2}
\providecommand{\BIBentrySTDinterwordspacing}{\spaceskip=0pt\relax}
\providecommand{\BIBentryALTinterwordstretchfactor}{4}
\providecommand{\BIBentryALTinterwordspacing}{\spaceskip=\fontdimen2\font plus
\BIBentryALTinterwordstretchfactor\fontdimen3\font minus
  \fontdimen4\font\relax}
\providecommand{\BIBforeignlanguage}[2]{{%
\expandafter\ifx\csname l@#1\endcsname\relax
\typeout{** WARNING: IEEEtran.bst: No hyphenation pattern has been}%
\typeout{** loaded for the language `#1'. Using the pattern for}%
\typeout{** the default language instead.}%
\else
\language=\csname l@#1\endcsname
\fi
#2}}
\providecommand{\BIBdecl}{\relax}
\BIBdecl

\bibitem{li2013spoken}
H.~Li, B.~Ma, and K.~A. Lee, ``Spoken language recognition: from fundamentals
  to practice,'' \emph{Proceedings of the IEEE}, vol. 101, no.~5, pp.
  1136--1159, 2013.

\bibitem{muthusamy1994reviewing}
Y.~K. Muthusamy, E.~Barnard, and R.~A. Cole, ``Reviewing automatic language
  identification,'' \emph{IEEE Signal Processing Magazine}, vol.~11, no.~4, pp.
  33--41, 1994.

\bibitem{nagarajan2004implicit}
T.~Nagarajan, ``Implicit systems for spoken language identification,'' 2004.

\bibitem{ambikairajah2011language}
E.~Ambikairajah, H.~Li, L.~Wang, B.~Yin, and V.~Sethu, ``Language
  identification: A tutorial,'' \emph{IEEE Circuits and Systems Magazine},
  vol.~11, no.~2, pp. 82--108, 2011.

\bibitem{mishra2023challenges}
J.~Mishra and S.~R.~M. {Prasanna}, ``Challenges in spoken language diarization
  in code-switched scenario,'' in \emph{2023 National Conference on
  Communications (NCC)}.\hskip 1em plus 0.5em minus 0.4em\relax IEEE, 2023, pp.
  1--6.

\bibitem{mishra2021spoken}
J.~Mishra, A.~Agarwal, and S.~R.~M. {Prasanna}, ``Spoken language diarization
  using an attention based neural network,'' in \emph{2021 National Conference
  on Communications (NCC)}.\hskip 1em plus 0.5em minus 0.4em\relax IEEE, 2021,
  pp. 1--6.

\bibitem{mishra2022issues}
J.~Mishra, J.~Gandra, V.~Patil, and S.~R.~M. {Prasanna}, ``Issues in
  sub-utterance level language identification in a code switched bilingual
  scenario,'' in \emph{2022 IEEE International Conference on Signal Processing
  and Communications (SPCOM)}.\hskip 1em plus 0.5em minus 0.4em\relax IEEE,
  2022, pp. 1--5.

\bibitem{dawalatabad2020novel}
N.~Dawalatabad, S.~Madikeri, C.~C. Sekhar, and H.~A. Murthy, ``Novel
  architectures for unsupervised information bottleneck based speaker
  diarization of meetings,'' \emph{IEEE/ACM Transactions on Audio, Speech, and
  Language Processing}, vol.~29, pp. 14--27, 2020.

\bibitem{sarma2018language}
M.~Sarma, K.~K. Sarma, and N.~K. Goel, ``Language recognition using time delay
  deep neural network,'' \emph{arXiv preprint arXiv:1804.05000}, 2018.

\bibitem{shah2020first}
S.~Shah, S.~Sitaram, and R.~Mehta, ``First workshop on speech processing for
  code-switching in multilingual communities: Shared task on code-switched
  spoken language identification,'' \emph{WSTCSMC 2020}, p.~24, 2020.

\bibitem{mishra2022importancelid}
J.~Mishra, S.~Siddhartha, and S.~M. Prasanna, ``Importance of excitation source
  and sequence learning towards spoken language identification task,'' in
  \emph{2022 National Conference on Communications (NCC)}.\hskip 1em plus 0.5em
  minus 0.4em\relax IEEE, 2022, pp. 190--194.

\bibitem{gupta2021clsril}
A.~Gupta, H.~S. Chadha, P.~Shah, N.~Chimmwal, A.~Dhuriya, R.~Gaur, and
  V.~Raghavan, ``Clsril-23: cross lingual speech representations for indic
  languages,'' \emph{arXiv preprint arXiv:2107.07402}, 2021.

\bibitem{liu2021end}
H.~Liu, L.~P.~G. Perera, X.~Zhang, J.~Dauwels, A.~W. Khong, S.~Khudanpur, and
  S.~J. Styles, ``End-to-end language diarization for bilingual code-switching
  speech,'' in \emph{22nd Annual Conference of the International Speech
  Communication Association, INTERSPEECH 2021}, vol.~2.\hskip 1em plus 0.5em
  minus 0.4em\relax International Speech Communication Association, 2021.

\bibitem{lyu2013language}
D.~C. Lyu, E.~S. Chng, and H.~Li, ``Language diarization for conversational
  code-switch speech with pronunciation dictionary adaptation,'' in
  \emph{Signal and Information Processing (ChinaSIP), 2013 IEEE China Summit
  and International Conference on}.\hskip 1em plus 0.5em minus 0.4em\relax
  IEEE, 2013, pp. 147--150.

\bibitem{barras2020vocapia}
C.~Barras, V.-B. Le, and J.-L. Gauvain, ``Vocapia-limsi system for 2020 shared
  task on code-switched spoken language identification,'' in \emph{The First
  Workshop on Speech Technologies for Code-Switching in Multilingual
  Communities}, 2020.

\bibitem{rallabandi2020detecting}
A.~S.~K. Rallabandi and A.~W. Black, ``On detecting code mixing in speech using
  discrete latent representations,'' \emph{WSTCSMC 2020}, p.~42, 2020.

\bibitem{yilmaz2017language}
E.~Yilmaz, M.~McLaren, H.~van~den Heuvel, and D.~A. van Leeuwen, ``Language
  diarization for semi-supervised bilingual acoustic model training,'' in
  \emph{Automatic Speech Recognition and Understanding Workshop (ASRU), 2017
  IEEE}.\hskip 1em plus 0.5em minus 0.4em\relax IEEE, 2017, pp. 91--96.

\bibitem{rangan2020exploiting}
P.~Rangan, S.~Teki, and H.~Misra, ``Exploiting spectral augmentation for
  code-switched spoken language identification,'' \emph{arXiv preprint
  arXiv:2010.07130}, 2020.

\bibitem{krishna2020utterance}
D.~Krishna and A.~Patil, ``Utterance-level code-switching identification using
  transformer network,'' \emph{WSTCSMC 2020}, p.~53, 2020.

\bibitem{park2022review}
T.~J. Park, N.~Kanda, D.~Dimitriadis, K.~J. Han, S.~Watanabe, and S.~Narayanan,
  ``A review of speaker diarization: Recent advances with deep learning,''
  \emph{Computer Speech \& Language}, vol.~72, p. 101317, 2022.

\bibitem{tranter2006overview}
S.~E. Tranter and D.~A. Reynolds, ``An overview of automatic speaker
  diarization systems,'' \emph{IEEE Transactions on audio, speech, and language
  processing}, vol.~14, no.~5, pp. 1557--1565, 2006.

\bibitem{moattar2012review}
M.~H. Moattar and M.~M. Homayounpour, ``A review on speaker diarization systems
  and approaches,'' \emph{Speech Communication}, vol.~54, no.~10, pp.
  1065--1103, 2012.

\bibitem{zhang2019fully}
A.~Zhang, Q.~Wang, Z.~Zhu, J.~Paisley, and C.~Wang, ``Fully supervised speaker
  diarization,'' in \emph{ICASSP 2019-2019 IEEE International Conference on
  Acoustics, Speech and Signal Processing (ICASSP)}.\hskip 1em plus 0.5em minus
  0.4em\relax IEEE, 2019, pp. 6301--6305.

\bibitem{fujita2019endpermuta}
Y.~Fujita, N.~Kanda, S.~Horiguchi, K.~Nagamatsu, and S.~Watanabe, ``End-to-end
  neural speaker diarization with permutation-free objectives,'' \emph{arXiv
  preprint arXiv:1909.05952}, 2019.

\bibitem{fujita2019end}
Y.~Fujita, N.~Kanda, S.~Horiguchi, Y.~Xue, K.~Nagamatsu, and S.~Watanabe,
  ``End-to-end neural speaker diarization with self-attention,'' in \emph{2019
  IEEE Automatic Speech Recognition and Understanding Workshop (ASRU)}.\hskip
  1em plus 0.5em minus 0.4em\relax IEEE, 2019, pp. 296--303.

\bibitem{bengio2007greedy}
Y.~Bengio, P.~Lamblin, P.~Popovici, and H.~Larochelle, ``Greedy layer-wise
  training of deep networks, advances in neural information processing systems
  19, mitpress, cambridge, ma,'' 2007.

\bibitem{baby2016resources}
A.~Baby, A.~L. Thomas, N.~Nishanthi, T.~Consortium \emph{et~al.}, ``Resources
  for indian languages,'' in \emph{Proceedings of Text, Speech and Dialogue},
  2016.

\bibitem{ravanelli2021speechbrain}
M.~Ravanelli, T.~Parcollet, P.~Plantinga, A.~Rouhe, S.~Cornell, L.~Lugosch,
  C.~Subakan, N.~Dawalatabad, A.~Heba, J.~Zhong \emph{et~al.}, ``Speechbrain: A
  general-purpose speech toolkit,'' \emph{arXiv preprint arXiv:2106.04624},
  2021.

\bibitem{ryant2018first}
N.~Ryant, K.~Church, C.~Cieri, A.~Cristia, J.~Du, S.~Ganapathy, and
  M.~Liberman, ``First dihard challenge evaluation plan,'' \emph{2018, tech.
  Rep.}, 2018.

\bibitem{lu2002speaker}
L.~Lu and H.-J. Zhang, ``Speaker change detection and tracking in real-time
  news broadcasting analysis,'' in \emph{Proceedings of the tenth ACM
  international conference on Multimedia}, 2002, pp. 602--610.

\bibitem{murty2008epoch}
K.~S.~R. Murty and B.~Yegnanarayana, ``Epoch extraction from speech signals,''
  \emph{IEEE Transactions on Audio, Speech, and Language Processing}, vol.~16,
  no.~8, pp. 1602--1613, 2008.

\bibitem{mishra2022importance}
J.~Mishra and S.~R.~M. {Prasanna}, ``Importance of supra-segmental information
  and self-supervised framework for spoken language diarization task,'' in
  \emph{International Conference on Speech and Computer}.\hskip 1em plus 0.5em
  minus 0.4em\relax Springer, 2022, pp. 494--507.

\bibitem{baevski2020wav2vec}
A.~Baevski, Y.~Zhou, A.~Mohamed, and M.~Auli, ``wav2vec 2.0: A framework for
  self-supervised learning of speech representations,'' \emph{Advances in
  Neural Information Processing Systems}, vol.~33, pp. 12\,449--12\,460, 2020.

\bibitem{bai2021speaker}
Z.~Bai and X.-L. Zhang, ``Speaker recognition based on deep learning: An
  overview,'' \emph{Neural Networks}, vol. 140, pp. 65--99, 2021.

\bibitem{mishra2023spoken}
J.~Mishra and S.~Prasanna, ``Spoken language change detection inspired by
  speaker change detection,'' \emph{arXiv preprint arXiv:2302.05265}, 2023.

\end{thebibliography}

\section{Biography Section}
 
\begin{IEEEbiography}
[{\includegraphics[width=1in,height=1.25in,clip,keepaspectratio]{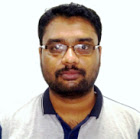}}]{Jagabandhu Mishra}
Jagabandhu Mishra (Student Member, IEEE) received the B. Tech. degree in Electronics and Telecommunication Engineering from BPUT Odisha, India, in the year 2015. He received the M. Tech. degree in Communication Engineering from the National Institute of Technology Nagaland, India, in the year 2018. He is currently pursuing a Ph. D. degree in Speech Signal Processing from the Dept. of Electrical Electronics and Communication Engineering, Indian Institute of Technology Dharwad, India. His research interests include speaker and language recognition, speech recognition, and speech signal analysis.
\end{IEEEbiography}

\begin{IEEEbiography}
[{\includegraphics[width=1in,height=1.25in,clip,keepaspectratio]{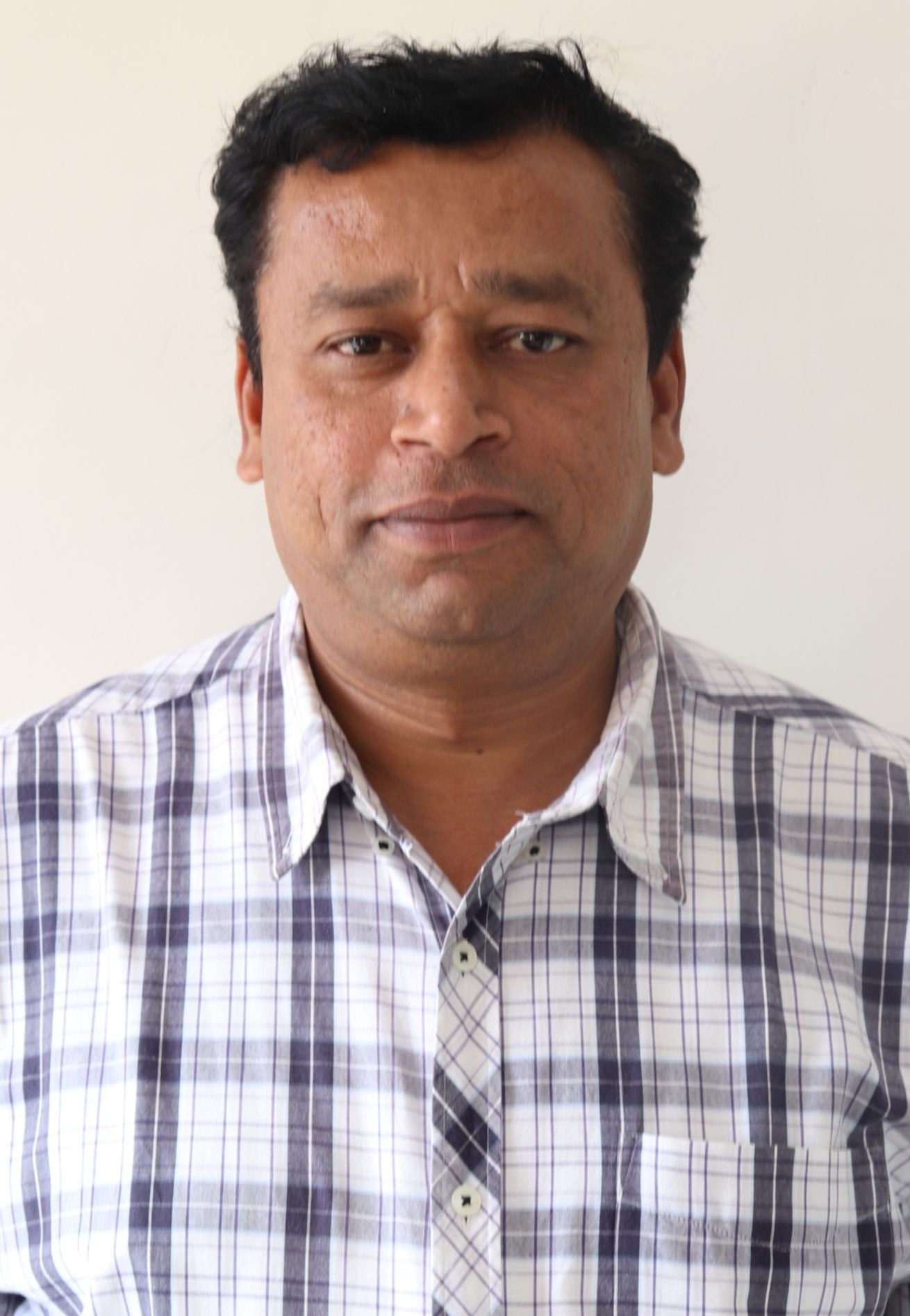}}]{S. R. Mahadeva Prasanna}
S. R. M. Prasanna (Senior Member, IEEE) is a Professor in the Department of Electrical Electronics and Communication Engineering, IIT Dharwad, Karnataka, India. He was earlier a Professor, Associate Professor, and Assistant Professor, all in the Dept. of Electronics and Electrical Engineering, IIT Guwahati, Assam, India. His areas of research interest include speech and handwriting processing, and applications of artificial intelligence, machine learning, and deep learning. He has supervised several Ph.D. Theses in these areas and published in reputed national and international journals and conferences.
\end{IEEEbiography}


\vspace{11pt}


\vfill

\end{document}